\documentclass{emulateapj_ed}

\newcommand{\kepler}{{\it Kepler}}
\newcommand{\mearth}{{\it MEarth}}

\usepackage{graphicx}
\usepackage{amsmath}
\usepackage{color}
\usepackage{natbib}

\usepackage{capt-of}

\shorttitle{Catalog of Stellar Rotation from the Kepler Data}
\shortauthors{McQuillan, Mazeh \& Aigrain}

\begin{document}
\title{Rotation Periods of 34,030 Kepler Main-Sequence Stars: The Full Autocorrelation Sample}

\author{A. McQuillan, T. Mazeh} 
\affil{School of Physics and Astronomy, Raymond and Beverly Sackler, Faculty of Exact Sciences, Tel Aviv University, 69978, Tel Aviv, Israel}
\email{amy@wise.tau.ac.il}
\author{S. Aigrain}
\affil{Department of Physics, University of Oxford, Oxford, OX1 3RH, UK\\}

\begin{abstract}
We analyzed 3 years of data from the \kepler\ space mission to derive rotation periods of main-sequence stars below 6500\,K. Our automated autocorrelation-based method detected rotation periods between 0.2 and 70 days for 34,030 (25.6\%) of the 133,030 main-sequence \kepler\ targets (excluding known eclipsing binaries and Kepler Objects of Interest), making this the largest sample of stellar rotation periods to date. In this paper we consider the detailed features of the now well-populated period-temperature distribution and demonstrate that the period bimodality, first seen by McQuillan, Aigrain \& Mazeh (2013) in the M-dwarf sample, persists to higher masses, becoming less visible above 0.6\,$M_\odot$. We show that these results are globally consistent with the existing ground-based rotation-period data and find that the upper envelope of the period distribution is broadly consistent with a gyrochronological age of 4.5\,Gyrs, based on the isochrones of \citet{bar07}, \cite{mam+08} and \cite{mei+09}. We also performed a detailed comparison of our results to those of  \cite{rei+13} and \cite{nie+13}, who have measured rotation periods of field stars observed by \kepler. We examined the amplitude of periodic variability for the stars with detected rotation periods, and found a typical range between $\sim 950$\,ppm ($5^{\rm th}$ percentile) and $\sim 22,700$\,ppm ($95^{\rm th}$ percentile), with a median of $\sim 5,600$\,ppm. We found typically higher amplitudes for shorter periods and lower effective temperatures, with an excess of low-amplitude stars above $\sim 5400$\,K.
\end{abstract}


\section{Introduction}

Throughout their main-sequence lifetime, stars lose angular momentum and spin down. For intermediate- and low-mass stars, this angular momentum loss is thought to occur via a magnetized wind that is linked to the stellar outer convection zone \citep[e.g.,][]{kaw88, bou+97}. Therefore, the present-day rotation period reflects the integrated angular momentum loss history of the star.

It has been claimed that for main-sequence stars, the present day rotation period does not depend on the initial conditions, and therefore a tight relationship exists between the period, age and mass, which is known as gyrochronology, for which a range of empirical relations have been derived \citep[e.g.,][]{kaw89, bar03, bar07, bar10, mam+08}. In recent years there has also been considerable work towards understanding and modeling the underlying physics behind gyrochronology \cite[][and references therein]{rei+12, mat+12, gal+13}. Therefore, deriving the rotation period for a large number of main-sequence stars has been a long-standing goal in stellar astrophysics, with the potential to shed light on stellar, and even planetary system angular-momentum evolution on one hand, and provide a new method to probe Galactic star formation history and structure on the other hand.

In recent years, the study of stellar rotation has seen significant advances thanks to new observational programs and techniques. Previously, measurements of stellar rotation were typically performed using spectroscopy, via the rotational broadening of absorption lines \cite[e.g.,][]{kal89}. However, this technique yielded only model-dependent constraints on the rotation rate due to the dependence on stellar radius and inclination, and were limited to relatively fast rotators.

The era of exoplanet transit surveys provided a new approach to stellar rotation studies, making it possible to measure the rotation period directly for a large number of stars simultaneously. This is done by detecting quasi-periodic brightness variations in the time series photometry, caused by magnetically active regions repeatedly crossing the visible hemisphere as the star rotates \citep[e.g.,][]{irw+09a}.

Open cluster surveys over the past decade have provided thousands of rotation periods for low-mass stars with ages up to $\sim 650$\,Myr, an overview of which can be found in \cite{irw+09} and \cite{bou+13}. More recent additions to the literature on rotation of young low-mass stars include studies of M35 (310 periods) \citep{mei+09},   Coma Berenices \citep{col+09}, M34 (83 periods) \citep{mei+11}, and h Persei (586) \cite{mor+13}. Together, these data have provided a relatively complete, but complex picture of pre-main-sequence rotational evolution. 

Until recently, rotation periods for older stars remained scarce, since their slow rotation and low-amplitude modulation make them difficult to detect with either spectroscopy or ground-based photometry. Notable exceptions include $112$ main sequence F, G and K stars observed as part of the Mount Wilson H-K project \citep[][and references therein]{bal+96}, $1727$ mid-F to mid-K stars observed by the {\it CoRoT}\ satellite \citep{aff+12}, 41 low-mass ($0.1$--$0.3\,M_\odot$) stars from the \mearth\ survey \citep{irw+11}, and $\sim 2000$ field K and M stars observed by the HATNet survey \citep{har+11}. The periods of \cite{aff+12} are based on up to 5 months of CoRoT data and yet include a significant fraction of periods greater than 60 days. Since many of these periods are more than half the length of the dataset, they should be treated with caution and we opted to omit these results from the comparison section of this paper.

Asteroseismology studies performed on a small subset of \kepler\ and CoRoT targets also provide an insight into stellar rotation, since the oscillation spectrum of a star can reveal information about surface and interior rotation rates \citep[e.g.,][]{giz02, giz+04}. For a recent review on observational studies of stellar rotation, see \cite{bou+13}.

Data from the Kepler space mission \citep{bor+10, koc+10} are revolutionizing the study of stellar rotation, providing 4\,yrs of almost uninterrupted photometry with an unprecedented level of precision and time sampling, for a very large sample of stars. This allows rotation periods to be detected for slowly rotating stars with low-amplitude modulation, for a well-defined large sample. 

Rotation studies on subsets of the \kepler\ data include the measurement of 265 rotation periods for stars with $T_{\rm eff} \le 5200\,$\,K and $\log g \ge  4.0$\,dex observed by \kepler\ for 1--2 quarters through the Cycle 1 Guest Observer program \citep{har+12}, 1570 M-dwarf rotation periods using quarters 1--4 (Q1--4) \citep{mcq+13a}, 737 of the Kepler Objects of Interest (KOIs) using Q3--14 data \citep{mcq+13b}, $\sim 950$ of the KOIs using Q9 \citep{wal+13}.

Two previous studies focussing on the broader \kepler\ sample are those of \cite{rei+13}, with an emphasis on differential rotation, who derive $\sim 24,000$ periods using Q3, and \cite{nie+13}, who measured $\sim 12,000$ periods from Q2--Q9, and compare to previous spectroscopic studies. These studies, discussed in Section~\ref{sec:disc}, use Fourier-based detection methods and conservative automatic selection criteria.

In this paper we describe the design and application of AutoACF, an automated version of the autocorrelation function (ACF) method introduced by \cite{mcq+13a} for stellar rotation period measurement. Using AutoACF we derived 34,030 rotation periods from 133,030 main-sequence targets observed by  \kepler. We outline the sample selection in Section~\ref{sec:data} and the period detection method in Section~\ref{sec:per_det}. In Section~\ref{sec:res} we present our results, and examine the mass-period and temperature-period distributions in Section~\ref{sec:per_mas_teff}, and the amplitude of periodic variability in Section~\ref{sec:amp_per}. In Section~\ref{sec:gyro} we compare our results to empirical gyrochronology relations, and in Section~\ref{sec:comp} we compare our results to those of \cite{rei+13} and \cite{nie+13}. We summarize our results in Section~\ref{sec:disc} and provide full details of the ACF automation method in the Appendix.

\section{Sample Selection}
\label{sec:data}

This analysis made use of the public releases 14--19 of quarter 3--14 (Q3--Q14) light curves (LCs), which were downloaded from the Kepler mission archive\footnote{http://archive.stsci.edu/kepler}. The data of Q0 and Q1 were omitted due to their short duration, and of Q2 due to significant residual systematics. We used the data corrected for instrumental systematics using PDC-MAP \citep{smi+12, stu+12}, which removes the majority of instrumental glitches and systematic trends using a Bayesian approach, while retaining most real variability.

The initial list of all targets observed by \kepler\ contains $\sim195,000$ stars. To select only main-sequence stars, we used the $T_{\rm eff}$--$\log g$ and color--color cuts advocated by \cite{cia+11}, which remove \mbox{$\sim 32,000$} giants from the \kepler\ sample. At this stage we also excluded $\sim 2000$ targets without $T_{\rm eff}$ or $\log g$ values. Where available for the low-mass stars, we use the improved values of \cite{dre+13} in place of the KIC parameters. A more recent version of the KIC is available \citep{hub+13}, however, we have opted to use the original since this provides the most homogeneous parameters. We checked that using the updated KIC values does not alter the results of this paper. 

Since eclipsing binary (EB) and planetary transit signals may affect the detection of rotation periods, we removed the 2611 known EBs and 4799 KOIs from the sample. The EB list was downloaded from the Villanova Eclipsing Binary web site\footnote{http://keplerebs.villanova.edu} which has an up-to-date and extended version of the \kepler\ EB catalogs of \cite{prs+11} and \cite{sla+11}. The EB catalog used in this work was downloaded on $7^{\rm th}$ Sept 2013. The KOI list was downloaded from the NASA Exoplanet Archive\footnote{http://exoplanetarchive.ipac.caltech.edu} \citep{ake+13} on $8^{th}$ Sept 2013 and includes confirmed planets, candidates, false positives and targets which have not been assigned a classification (labelled `not dispositioned').

At this stage 154,925 targets remained in our list. Some targets were not observed in all quarters, and hence we selected only targets observed in at least 8 out of the 12 quarters used in this study, leaving a sample of 139,638 stars. Finally, since we are interested only in the rotation periods of stars with convective envelopes which spin down during their lifetimes, we introduce a limit of $T_{\rm eff} < 6500$\,K. This also greatly reduces the potential for contamination of the sample by pulsating stars from the instability strip. The final sample on which the period search was run consisted of 133,030 stars.\\

\section{Period Detection}
\label{sec:per_det}

Periodic detection was performed using the ACF method, described in detail in McQuillan, Aigrain \& Mazeh (2013). The ACF measures the degree of self-similarity of an LC over a range of time lags. In the case of rotational modulation, repeated spot crossings over the visible stellar hemisphere produce ACF peaks at lags corresponding to the stellar rotation period and its integer multiples. 

The ACF has been shown to produce clear and robust results even when the amplitude and phase of the photometric modulation evolve significantly, and when systematic effects and long-term trends are present \citep{mcq+13a}. For this reason, we adopt the ACF method of period detection over the more widely used Fourier-based methods, which can be strongly affected by residual systematics. The clarity of output and the robust nature of the ACF make it amenable to automation, without requiring such conservative thresholds as the Fourier-based approaches. See \cite{mcq+13a} for a detailed comparison of the ACF method with Fourier-based approaches.

We preprocessed the LCs following the method of \cite{mcq+13a}, in which we median normalized each $\sim 90$ day quarter of \kepler\ data. In an update to the first application of this code, we now map the data to an evenly spaced grid of 29.4 minutes to match the \kepler\ long cadence, and fill any missing flux values (for example in the gaps between quarters) with zeros to prevent them contributing to the ACF.

We then computed the ACF for each LC and identified the peak corresponding to the rotation period, which is followed by series of peaks at lags equal to integer multiples of the rotation period. We then used the locations of this set of peaks to determine the exact period of the target LC. In a minor modification to the original code, we defined the period as the gradient of a straight-line fit to the ACF peak positions, as a function of peak number. Up to 4 consecutive peaks at near integer multiples of the selected ACF peak were used, as well as the intercept at (0,0). This provides a more reliable period measurement and uncertainty than the median and scatter of the peaks, which were used by \cite{mcq+13a}.

For stars where active regions are present on opposite hemispheres of the star, a partial correlation exists at half the rotation period, leading to additional peaks in the ACF. If the second peak in the ACF is larger than those on either side, the ACF code identifies the larger peak (and integer multiples of it) as presenting the correct rotation period. Of the 5567 cases for which the ACF selected the second peak, we visually checked all cases for which the peak height difference between the first and the second peak was less than 20\%, and corrected any erroneous peak selections.

\cite{mcq+13a} focussed on a small sample of 2483 cool \kepler\ targets, and visually examined all LCs to verify each rotation period detection by the ACF. Since this is not feasible for the 133,030 stars in our sample, we have automated the selection procedure (AutoACF). Using a visually verified training set with 1000 confirmed periodic targets per \mbox{500\,K} $T_{\rm eff}$ bin, we sought the parameters of the ACF which separate likely true period detections from false positives originating from noise or instrumental systematics. 

We performed the selection in two stages, first testing the consistency of the ACF detection in separate data sections, and then the significance of the detection. A good criterion for real astrophysical periodicity is that it can be detected in multiple regions of the LC, since ACF peaks caused by systematics or artifacts are less likely to appear in different section of the LC. Therefore, we required the period found for the whole Q3--Q14 LC to be detected also in 2 or more individual segments, where a segment is defined as 3 consecutive \kepler\ quarters (Q3--Q5, Q6--Q8, Q9--Q11, Q12--Q14). 

To identify real periods based on the detection significance, we used the height of the primary ACF peak with respect to the troughs on each side, which we call `local peak height' (LPH). Clear period detections have higher LPHs than those suspected to originate from noise. We assigned a weighting, $w$, to each period detection according to its LPH and position in $T_{\rm eff}$-LPH-Period space. A threshold $w_{\rm thres}$, can then be selected to obtain the desired compromise between real detections and false positives. For this work we selected $w_{\rm thres} = 0.25$. We tested that altering the input parameters and functions used in AutoACF, and the threshold weight, $w_{\rm thres}$, does not significantly alter the resulting period distribution. Full details of the selection method are given in Appendix~\ref{sec:auto}.

The first application of AutoACF derived rotation periods for 33,694 stars. We then compared our results to those of \cite{rei+13} and \cite{nie+13} who have also performed rotation period studies of the \kepler\ sample. This comparison, described in detail in Section~\ref{sec:comp_sub}, identified a subset of stars for which both other works were able to detect rotation periods and AutoACF was not. These missed detections had two dominant origins: faults in the LC, and multiple periods. We implemented an additional level of processing to the LCs and ACFs of stars where no rotation period had been determined to detect periods initially missed due to faults (marked as `BQR' in Table~\ref{tab:main_res}) and multiple periods (marked as `SM1' and `SM2' in Table~\ref{tab:main_res}). Details of this procedure are described in Appendix~\ref{sec:add}. These additions bring the total number of rotation period detections to 34,030. Differences between the results of \cite{mcq+13a} and the current study are minimal and dominated by the improvement in detection from the longer baseline, rather than the difference between the methods.

\begin{table*}
  \caption{The rotation period measurements for the 34,030 stars presented in this work. This table is available in its entirety in a machine-readable form in the online supplementary material and at {\tt www.astro.tau.ac.il/$\sim$amy}. A portion is shown here for guidance regarding its form and content. $T_{\rm eff}$ and $\log g$ are from the KIC or, where available, from \cite{dre+13}, indicated by a DC value of 1. The mass, $M$, was derived from $T_{\rm eff}$ using the isochrones of \cite{bar+98}. The remaining columns are: period, $P_{\rm rot}$, and period error, $\sigma_{\rm P}$; the average amplitude of variability within one period, $R_{\rm per}$; the local peak height, LPH; and the assigned weight, $w$. The Flag column indicates whether fault correction (`BQR'), soft smoothing (`SM1') or hard smoothing (`SM2') were required to detect a period (see Appendix~\ref{sec:auto} and~\ref{sec:add} for further details).}
  \centering
  \begin{tabular}{ccccccccccc} 
    \hline
     KIC &  $T_{\rm eff}$ & $\log g$ & $M$               & $P_{\rm rot}$ & $\sigma_{\rm P}$ & $R_{\rm per}$ & LPH & $w$ & DC & Flag \\
             &  (K)                  & (dex)      &  ($M_\odot$) & (days) & (days) & (ppm) & & &  &\\
    \hline  
     892376&3813&4.47&0.4699&1.532&0.007&7306.69&0.823&0.4503&0&---\\
1026146&4261&4.57&0.6472&14.891&0.120&11742.56&1.405&0.7229&0&---\\
1026474&4122&4.56&0.5914&1.569&0.006&30471.80&1.204&0.6061&0&---\\
1162635&3760&4.77&0.4497&15.678&0.019&10207.47&0.978&0.5445&1&---\\
1164102&4045&4.62&0.5606&31.496&0.474&5139.74&0.568&0.3939&0&---\\
    \hline
 \end{tabular}
 \label{tab:main_res}
\end{table*}

\begin{table*}
  \caption{Details of the 99,000 stars with no significant period detection. This table is available in its entirety, in a 
  machine-readable form in the online supplementary material and at {\tt www.astro.tau.ac.il/$\sim$amy}. A portion 
  is shown here for guidance regarding its form and content. 
  Column descriptions are the same as for Table~\ref{tab:main_res}. 
  Targets without a $w$ value were rejected at selection process stage 1 
  because the period detection did not occur in enough segments 
  (see Appendix~\ref{sec:auto} for details). In these cases, $P_{\rm rot}$, $\sigma_{\rm P}$, LPH and $w$ are
  marked as `nan'.}
  \centering
  \begin{tabular}{cccccccccc} 
    \hline
     KIC &  $T_{\rm eff}$ & $\log g$ & $M$               & $P_{\rm rot}$ & $\sigma_{\rm P}$  & LPH & $w$ & DC \\
             &  (K)                  & (dex)      &  ($M_\odot$) & (days) & (days)  & & & \\
    \hline  
    	893305&4133&4.58&0.5958&nan&nan&nan&nan&0\\
	1027110&4155&4.50&0.6046&1.701&0.039&0.299&0.1439&0\\
	1027277&4326&4.57&0.6735&60.136&0.691&0.315&0.0876&0\\
	1160660&4232&4.59&0.6355&nan&nan&nan&nan&0\\
	1160684&3952&4.48&0.5239&0.419&0.090&0.150&0.0266&0\\
    \hline
 \end{tabular}
 \label{tab:rem_res}
\end{table*}

\begin{figure*}
  \centering
  \includegraphics[width=\linewidth]{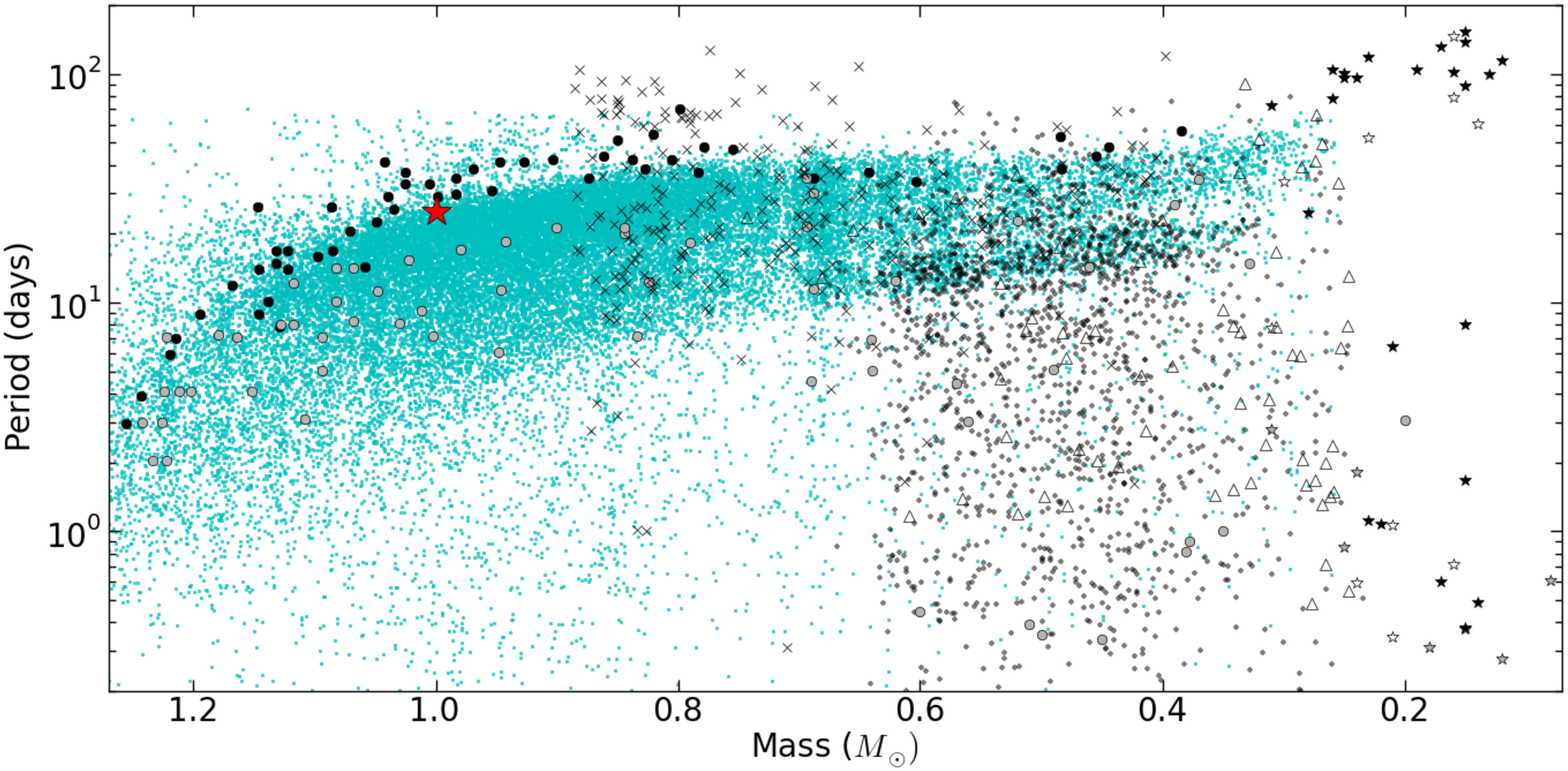}
  \caption{Period versus mass with comparison to previous rotation period measurements.
    The 34,030 new rotation periods derived using AutoACF are shown as 
    cyan points. The mass was derived using the models of \cite{bar+98}, 
    as described in the text. This figure also displays periods 
    from \citet{bal+96} and \citet{kist07} (114 circles) and \mearth\ data from
    \citet{irw+11} (41 stars), with grey and black symbols representing
    objects with young and old disk kinematics,
    respectively, all of which have available mass estimates. Additional M-dwarf periods
    from the WFCAM Transit Survey \citep{gou+12}, for which no
    kinematic classification is available (65 triangles), with masses derived from \cite{mam11}. 
    Also included are periods from \cite{har+11} (1686 small black dots), with mass estimates 
    obtained using $T_{\rm eff}$ and the models of \cite{bar+98}, and
    periods from \cite{har+12} (265 crosses), with masses derived from
    a  $J-K$ to $T_{\rm eff}$ conversion using data from \cite{ken+95}, and the isochrones of \cite{bar+98}.}
  \label{fig:kep_mass_1}
\end{figure*}

\section{Results}
\label{sec:res}

Using AutoACF we derived rotation periods for 34,030 of the stars in the \kepler\ sample defined in Section~\ref{sec:data}. These results are shown in Table~\ref{tab:main_res}.

From the application of AutoACF to the visually examined training set, we estimate that $> 95$\% of the derived periods represent true rotation periods. Of the remaining $\sim 5$\%, half are probable rotation period detections and half are false detections triggered by systematics or likely pulsators (based on visual examination of the LC and ACF). Based on the application of AutoACF to the training sample, we estimate that an additional $\sim 5000$ periods could have been detected by eye, and that these are predominantly long-period or low-amplitude signals. It may be possible to augment the periodic sample by visual examination of targets falling just below the selected $w_{\rm thres}$ value.

Table~\ref{tab:rem_res} shows the subset of stars to which we applied AutoACF but the resulting detection did not meet our criteria. Table~\ref{tab:fracs} shows the periodic detection fraction per 500\,K $T_{\rm eff}$ bin. This fraction goes from $\sim 0.8$ for the coolest stars, with temperatures below 4000\,K, to $\sim 0.2$ around 6000\,K.

\begin{minipage}[t]{0.98\textwidth}
  \includegraphics[width=\linewidth]{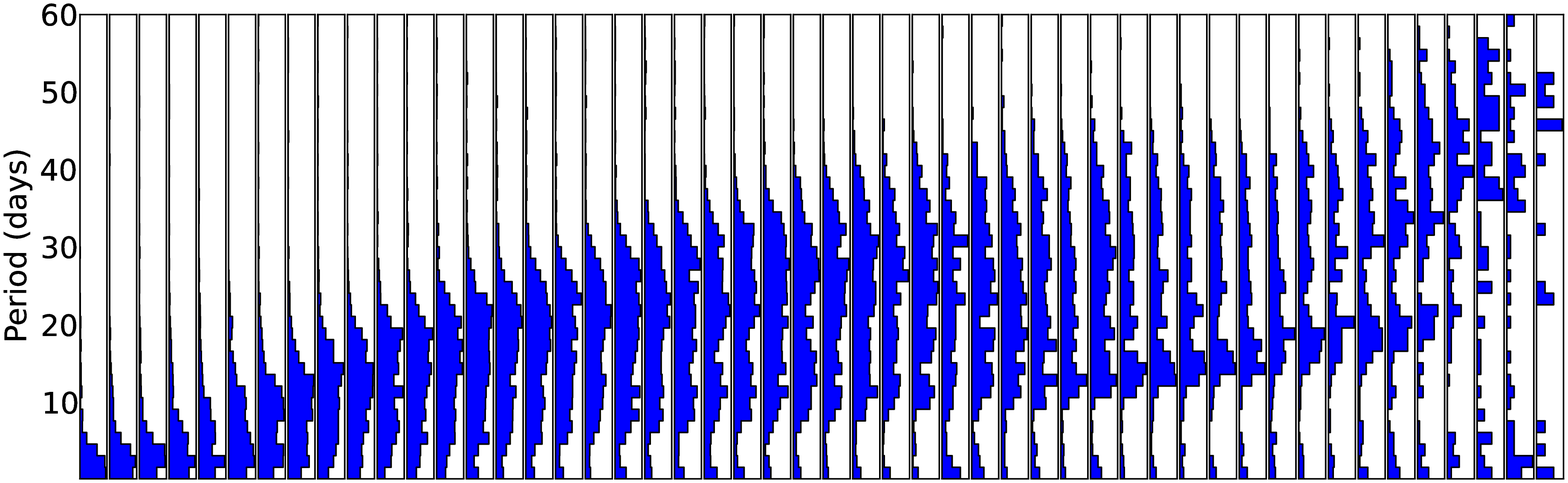}
  \includegraphics[width=\linewidth]{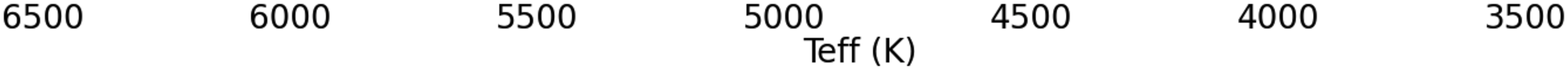}
  \captionof{figure}{Period versus $T_{\rm eff}$ shown in histogram form for each temperature bin, clearly showing the bimodality in rotation period for cooler stars, and the slight dip in the upper envelope of the period distribution at $\sim 4000$\,K.}
  \label{fig:hists}
\end{minipage}

\begin{table}[h!]
  \caption{Periodic fractions across the temperature range examined.}
  \centering
  \begin{tabular}{cc} 
    $T_{\rm eff}$ (K) & Periodic Fraction \\
    \hline
    $< 4000$ & 0.83\\
    4000--4500 & 0.69\\
    4500--5000 & 0.43\\
    5000--5500 & 0.27\\
    5500--6000 & 0.16\\
    6000--6500 & 0.20\\
 \end{tabular}
 \label{tab:fracs}
\end{table}

\subsection{Mass-Period and Temperature-Period Distributions}
\label{sec:per_mas_teff}

Figure~\ref{fig:kep_mass_1} shows the mass-period distribution of the 34,030 stars with measured rotation periods, together with period derivations from previous work, most of which originate from ground-based observations. Mass, $M$, is calculated from the KIC $T_{\rm eff}$ using the stellar evolution models of \cite{bar+98}, using isochrone no.\,1 for $M < 0.7M_{\odot}$ and isochrone no.\,3 for higher masses, and assuming an age of $\sim 1$\,Gyr. We checked that the change in results is negligible if the age is varied by a factor of up to 10. The typical uncertainty associated with the KIC $T_{\rm eff}$ values is 200\,K, which translates to a uncertainty in mass of somewhat less than $0.1\,M_\odot$. Vertical features in Figure~\ref{fig:kep_mass_1}, such as the gaps at $\sim 0.55\,M_\odot$ and $\sim 0.7\,M_\odot$, are artifacts introduced by the KIC temperature information and are not real. Conversion between $B-V$ and $T_{\rm eff}$ where required in this work was performed using the equations of \cite{sek+00}.

The period measurements presented in this work are consistent with the existing ground-based photometric rotation period data, showing a trend of typically increasing rotation period with decreasing mass. The Sun, marked in Figure~\ref{fig:kep_mass_1} as a red star, sits on the upper envelope period distribution.

The bimodality in period distribution, first reported by \cite{mcq+13a} for the M-dwarf sample, is clearly visible in the low-mass half of Figure~\ref{fig:kep_mass_1}. To explore the bimodality further, we plotted the data as a set of histograms, which are shown in Figure~\ref{fig:hists}. To eliminate conversion uncertainties between $T_{\rm eff}$ and mass, we plotted the $T_{\rm eff}$-period distribution in this figure. Each histogram is normalized, such that only frequencies on the period scale can be directly compared, \\\\\\\\\\\\\\\\\\\\\\\\\\\\\\\\\\\\\\\\ and not on the $T_{\rm eff}$ scale. 

This histogram representation increases the clarity of the period bimodality in the low-$T_{\rm eff}$ region, and the width of the gap between the two sequences can be seen to increase towards cooler temperatures. At $\sim 3500$\,K the two peaks are at $\sim 20$\,days and $\sim 40$\,days, with a Hartigan's dip test \citep{hart85} p-value for unimodality of 0.01. At $\sim 4000$\,K the two peaks are at $\sim 14$\,days and $\sim 30$\,days, with a Hartigan's dip test p-value for unimodality of 0.15. This bimodal sequence is not visible above $\sim4500$\,K. 

Figure~\ref{fig:hists} also shows that the upper envelope of periods increases steadily with decreasing temperature, from the hottest stars down to $\sim$~4500\,K, at which point the long-period envelope decrease slightly before rising again below $\sim 4000$\,K.

\subsection{Amplitude of Periodic Variability}
\label{sec:amp_per}

\begin{figure}[!h]
  \centering
  \includegraphics[width=\linewidth]{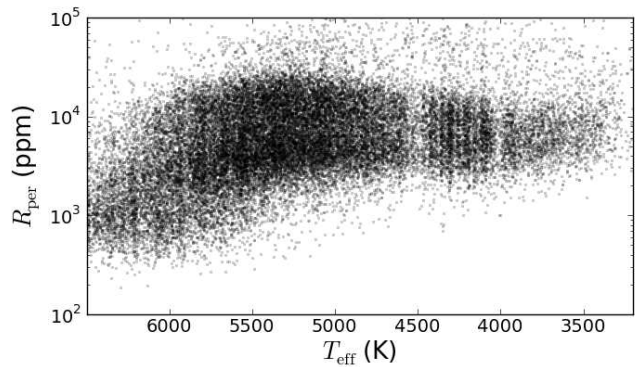}
  \caption{The distribution of periodic photometric variability amplitude, $R_{\rm per}$, over the temperature range for all stars with derived rotation periods.}
  \label{fig:teff_amp}
\end{figure}

\begin{figure*}
  \centering
  \includegraphics[width=0.325\linewidth]{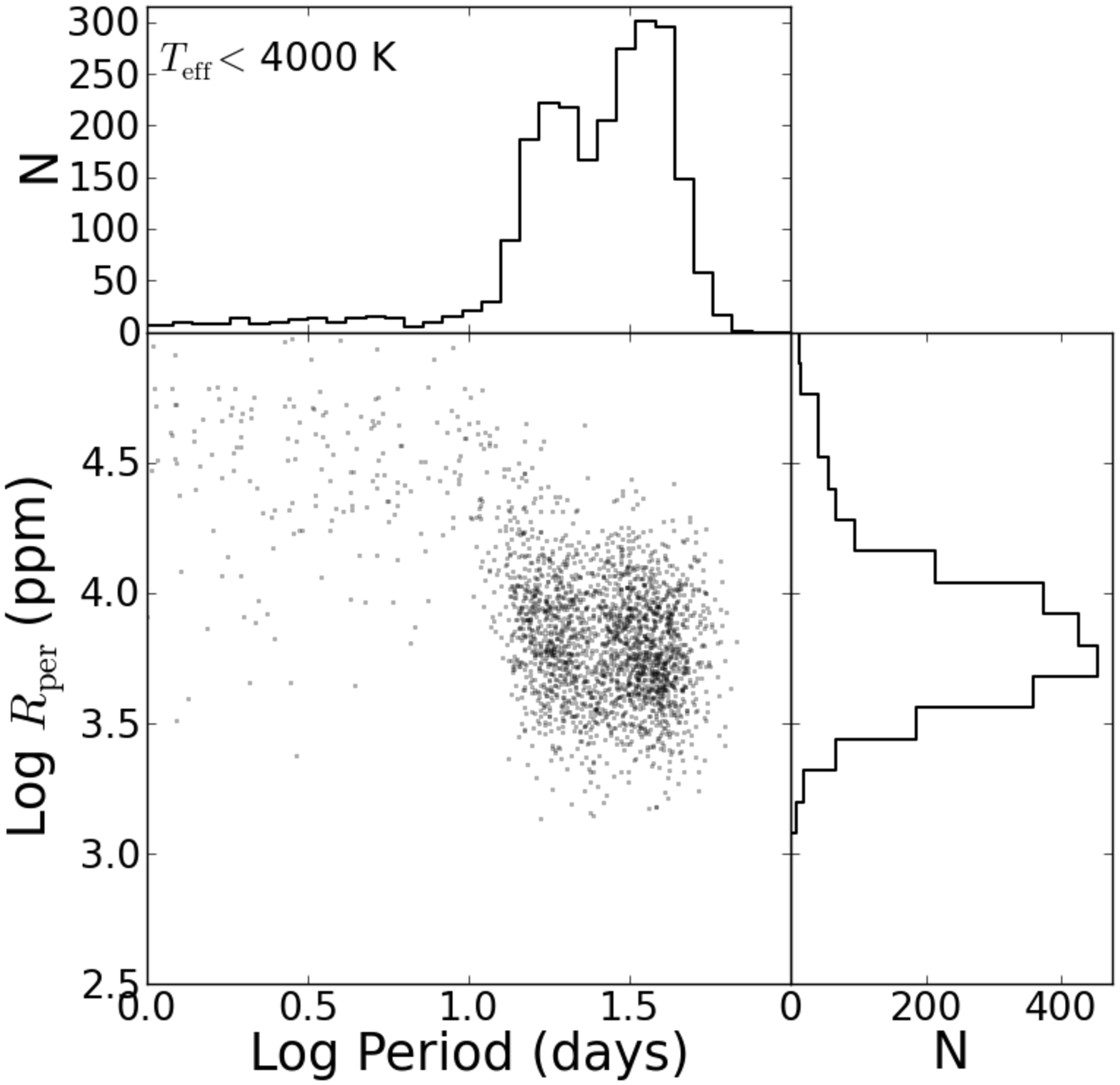}
  \includegraphics[width=0.325\linewidth]{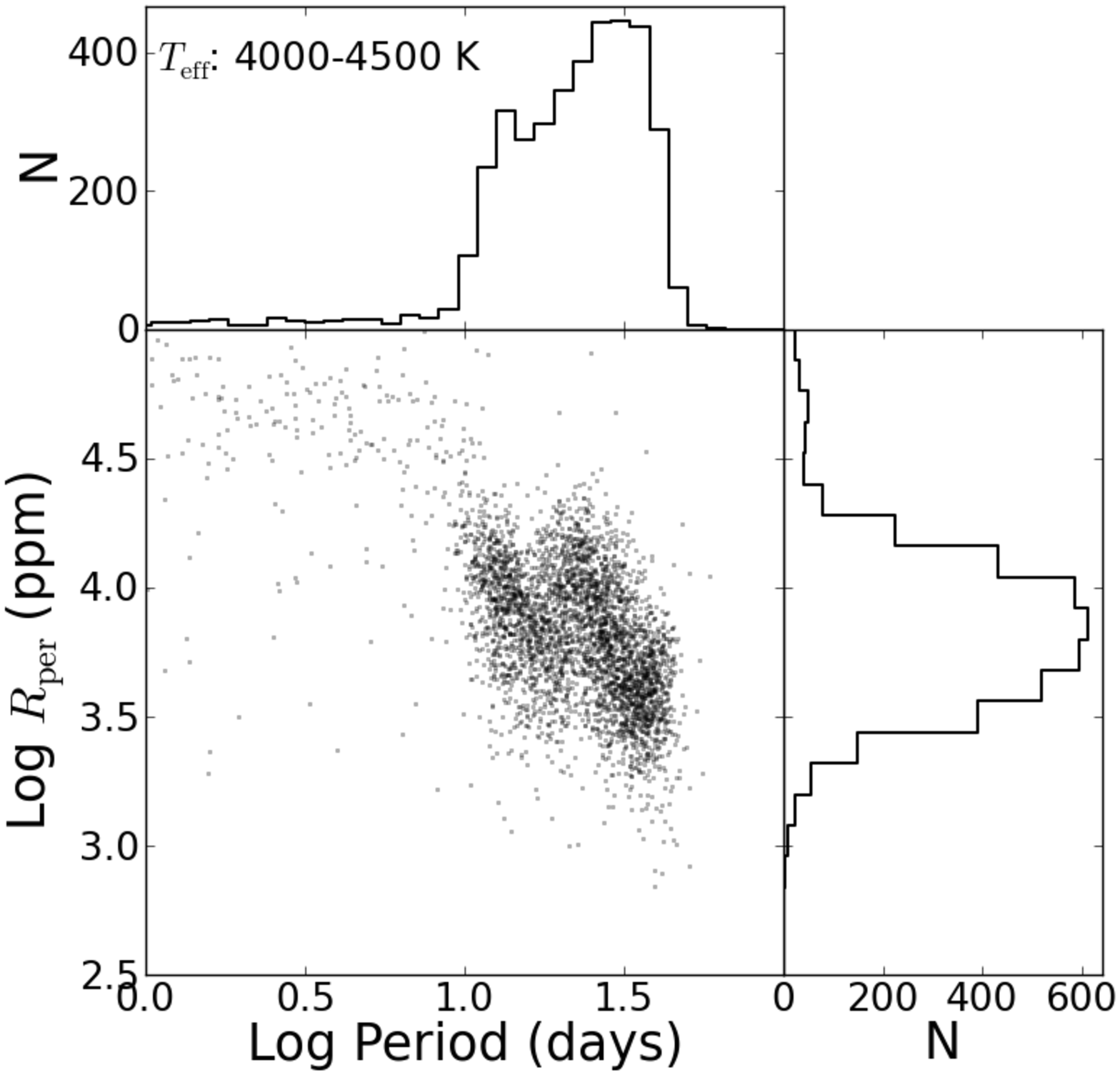}
  \includegraphics[width=0.325\linewidth]{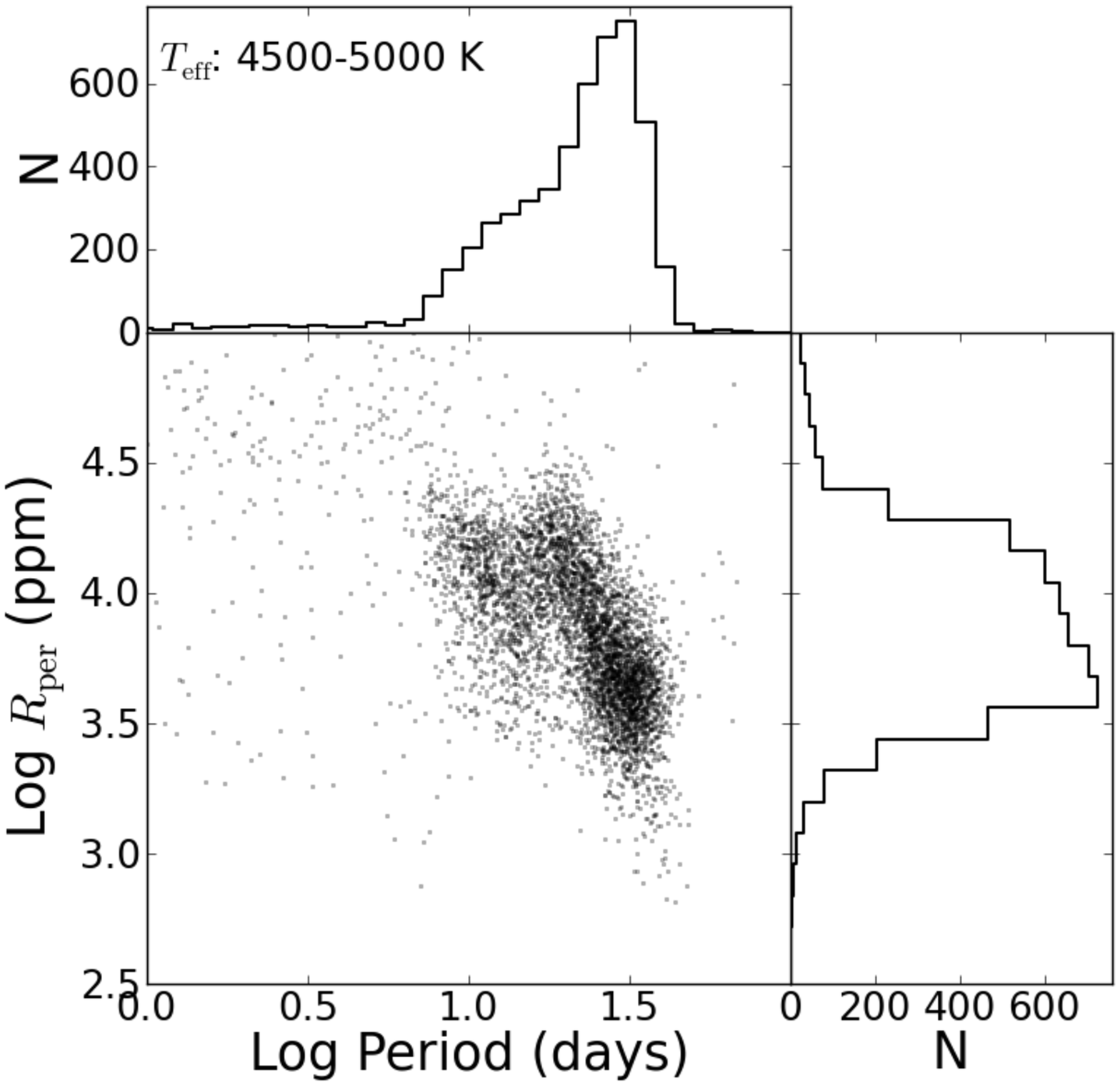}
  \includegraphics[width=0.325\linewidth]{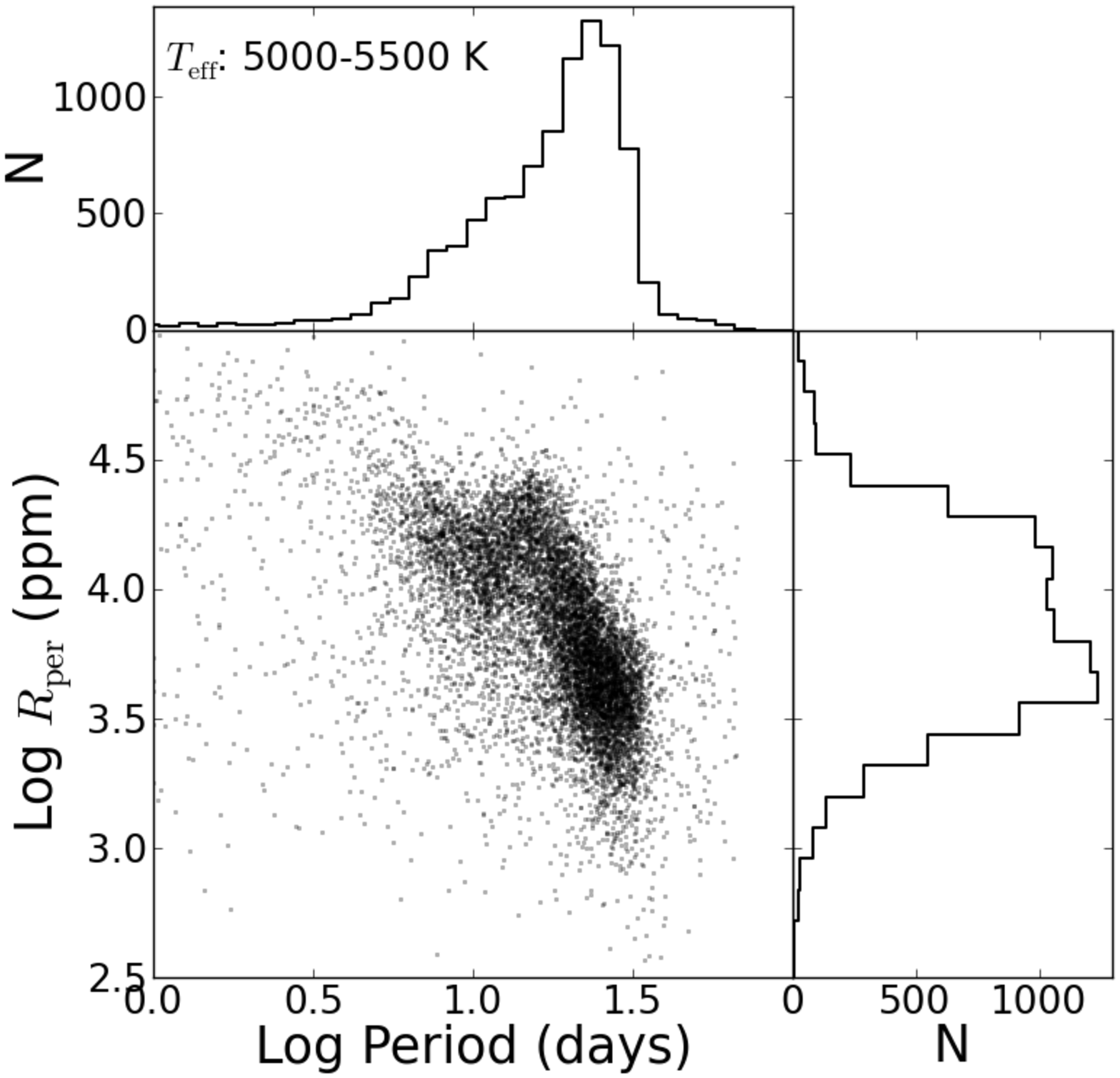}
  \includegraphics[width=0.325\linewidth]{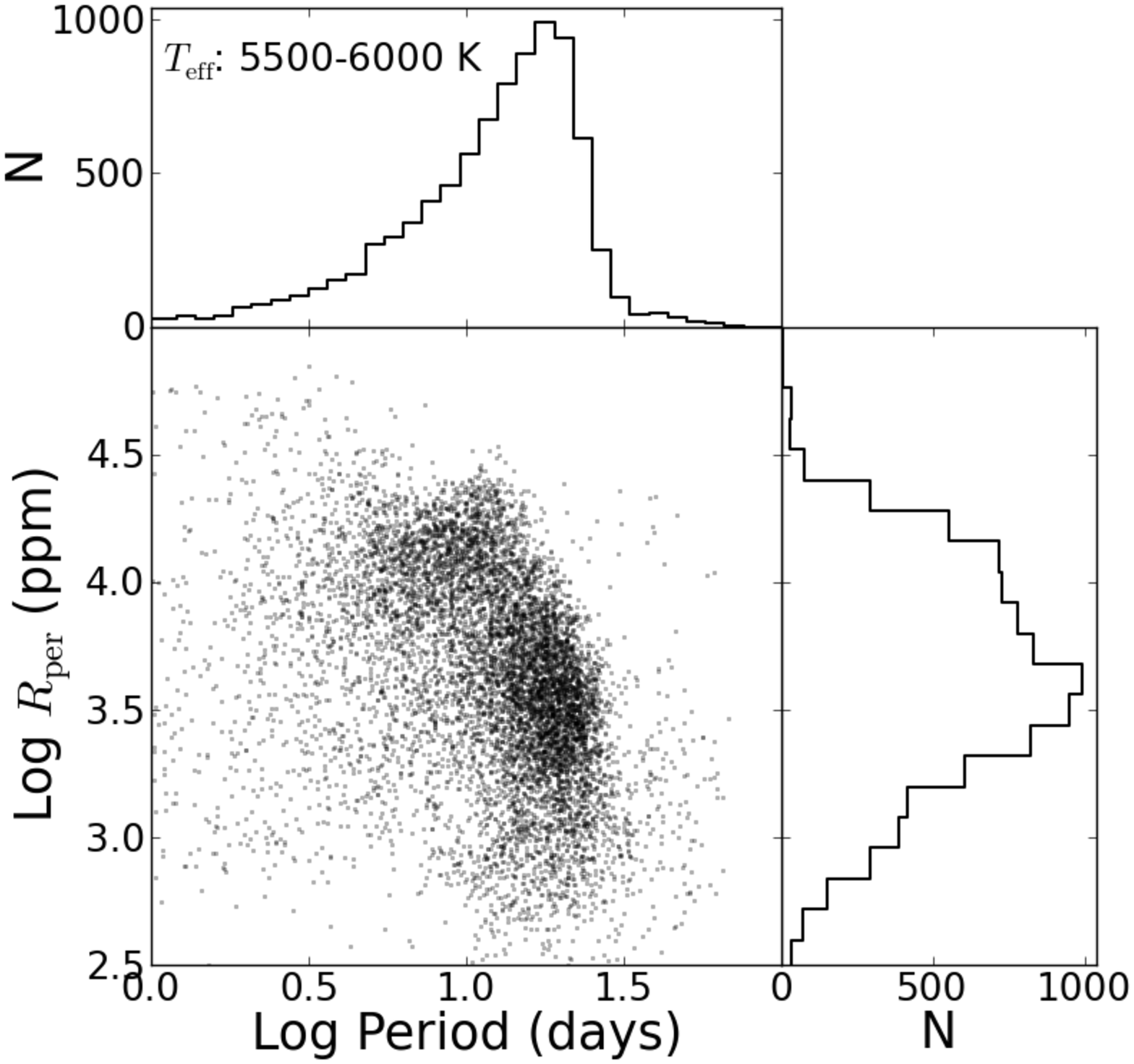}
  \includegraphics[width=0.325\linewidth]{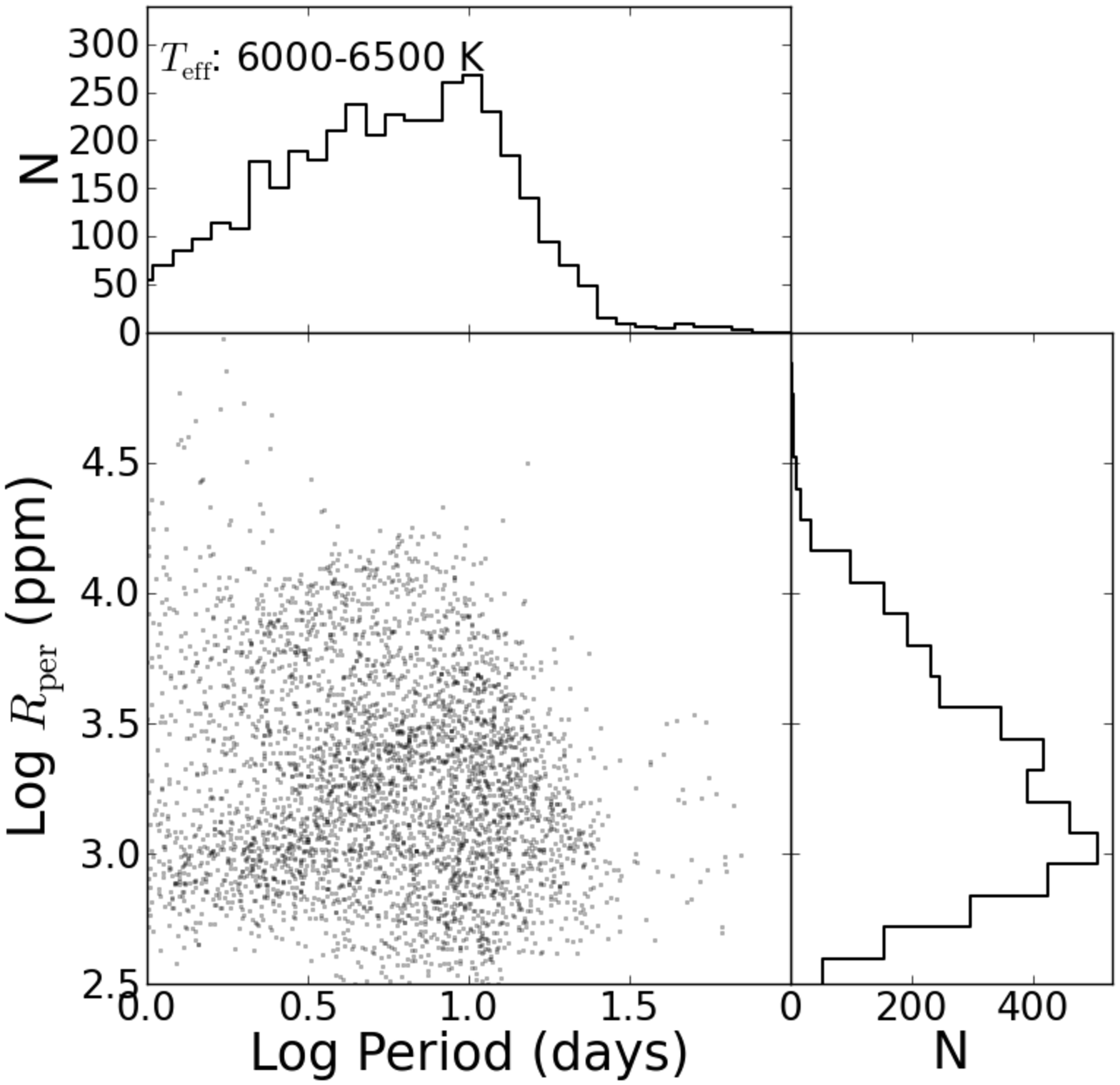}
  \caption{ Log amplitude of periodic variability, $R_{\rm per}$, against log period (for periods $>$ 1 day), for each 500\,K $T_{\rm eff}$ bin, as described in the top section of each panel. The top section of each panel shows the histogram of log period and the side section of each panel shows the histogram of log $R_{\rm per}$.}
  \label{fig:amp_per}
\end{figure*}

For each LC, we derived the amplitude of the photometric variability. We define this as a measure of the range between the $5^{\rm th}$ and $95^{\rm th}$ percentile of normalized flux, as described in \cite{bas+11} and \cite{mcq+13a}. To ensure we measure the amplitude of the periodic variability and not long-term trends in the data, we divide each LC into sections of period length and measure the $5^{\rm th}$ to $95^{\rm th}$ percentile of normalized flux in each section. We then take the median of these values as the amplitude of periodic variability, $R_{\rm per}$, for the LC. 

Figure~\ref{fig:teff_amp} shows $R_{\rm per}$ versus $T_{\rm eff}$ for the stars with detected rotation periods, and demonstrates a dearth of low-amplitude modulation at low temperatures. The range of amplitudes increases towards higher temperatures, with both an increase in the upper envelope of amplitudes and an excess of low-amplitude stars below $\sim 5400$\,K. The typical range of amplitudes is between $\sim 950$\,ppm ($5^{\rm th}$ percentile) and $\sim 22,700$\,ppm ($95^{\rm th}$ percentile), with a median of $\sim 5,600$\,ppm. We find typically higher amplitudes for shorter periods and lower effective temperatures.

We checked to see whether photon noise was artificially increasing the amplitude for fainter stars by examining the same distribution for only the brighter targets (Kepler magnitude $< 14$), and saw the same trend. The dearth of low-amplitude stars at low temperatures has been previously observed \citep{bas+10, mcq+12}, but this is the first study to isolate variability resulting from rotational modulation. Since this figure shows only the stars for which rotation periods were detected, it suffers from the same biases as the period detection, namely that the long-period, low-amplitude stars are hardest to detect.

We also examined the distribution of photometric variability as a function of period, which is shown across the range of temperatures in Figure~\ref{fig:amp_per}. Previous studies have suggested a typical increase in variability with rotation rate \citep[e.g.,][]{piz+03, har+11}, although this was not seen by \cite{bas+11}. In each $T_{\rm eff}$ bin a negative correlation between rotation period and $R_{\rm per}$ is evident, although the picture becomes somewhat more complex for the hottest stars. The cooler stars show a sequence running from high-amplitude, short-period stars, down to low-amplitude, long-period stars. This effect is unlikely to arise as a result of detection bias, since period detection is hardest for low amplitudes and long periods, and thus would not exclude the scarcely populated regions on either side of the sequence (low-amplitude and short-period, and high-amplitude and long-period).

The two period groups, shown clearly by the bimodality in the coolest $T_{\rm eff}$ bin, are visible up to 5500\,K. The period-amplitude gradient prevents the bimodality from appearing in the period histograms, except for in the coolest set where it is most prevalent. It is interesting to note that the amplitudes of the short- and long-period groups in the coolest stars have similar amplitudes, whereas slower rotators extend to lower amplitudes for the early K and G stars.

\section{Gyrochronology}
\label{sec:gyro}

\begin{figure*}
  \centering
  \includegraphics[width=0.9\linewidth]{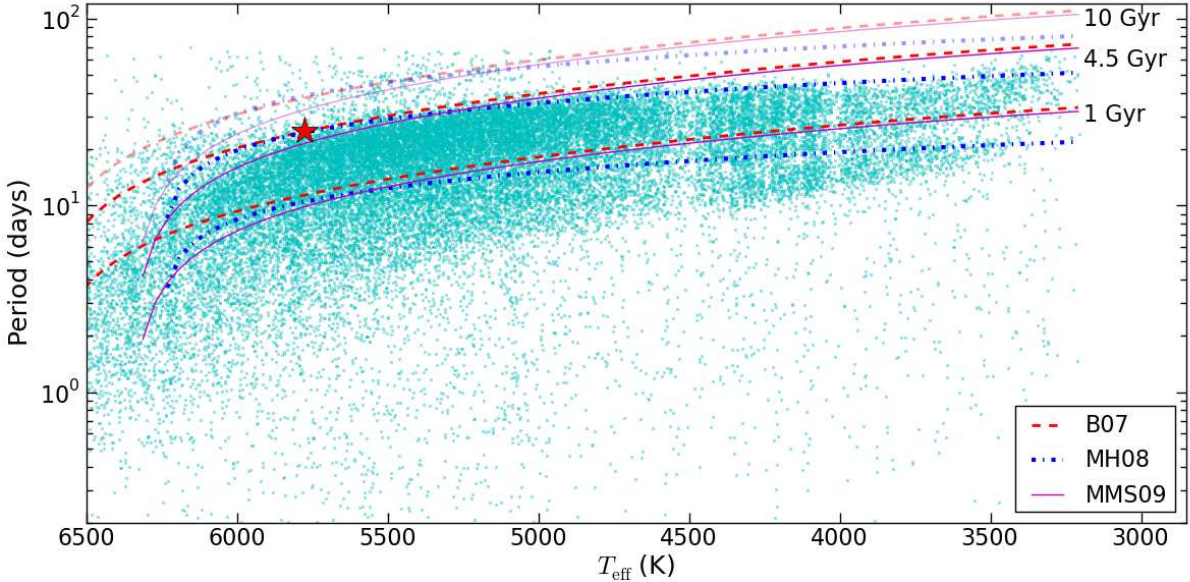}
  \caption{The $T_{\rm eff}$-period distribution showing the location of selected rotational isochrones. The red dashed line marks the 1\,Gyr (lower), 4.5\,Gyr (middle) and 10\,Gyr (upper, faint) empirical gyrochronology relation of \cite{bar07}, denoted B07. The blue dot-dashed line marks the equivalent age isochrones from the empirical gyrochronology relation of \cite{mam+08}, denoted MH08, and the solid magenta line marks that of \cite{mei+09}, denoted MMS09.}
  \label{fig:gyro}
\end{figure*}

In Figure~\ref{fig:gyro} we compare the $T_{\rm eff}$-period sequence to the commonly used empirical gyrochronology models of \cite{bar07}, \cite{mam+08} and \cite{mei+09}. We find that the upper envelope of the period distribution is broadly consistent with a gyrochronological age of 4.5\,Gyrs. The models all appear to under-predict stellar ages, since they imply a significant proportion of the stars are $< 1$\,Gyr and very few are $> 4.5$Gyr, which would be unlikely for a field star population such as that observed by \kepler. Although the spin-down rate reduces with age, stars older than 4.5\,Gyr should still appear distinctly above the upper envelope of points, as shown by the 10\,Gyr isochrone. However, the observational biases are not well understood and the oldest stars are the hardest to derive photometric rotation periods for since they have long rotation periods and low-amplitude variability.

None of the models capture the slight dip in the upper envelope of rotation periods around 3600--3900\,K, or the more pronounced increase in rotation periods for the M-dwarfs. None of the models are able to accurately reproduce the shape of the lower envelope of the distribution either, which should be less affected by observational or measurement biases than the upper envelope. 

There is also a subset of very rapid rotators across all temperatures with periods below $\sim 15$ days, implying very young ages based on the gyrochronology models. This sample of very short-period rotators is considerably larger than would be expected for the \kepler\ sample, which are predominantly field stars and should not have a large number of stars younger than 1\,Gyr. It is therefore unlikely that these objects are fast rotators due to youth. Possible alternative explanations include the hypothesis that these are non-eclipsing binaries, where spin-orbit interaction has lead to the rapid rotation speed, as suggested by \cite{mcq+13a}, contaminating pulsation measurements, or that the more massive of these stars may represent the sub-giant contamination of the \kepler\ sample \citep{vs+13}.

\section{Comparison with Previous Kepler Rotation Studies}
\label{sec:comp}

\begin{table*}[h!]
  \caption{A numerical comparison of AutoACF, \cite{rei+13} and \cite{nie+13}. The top row contains the 
  total number of reported rotation period measurements from each work. The next three rows show the number in both the column and row labelled work.}
  \centering
  \begin{tabular}{|l || c | c | c |} 
    \hline
                                  & AutoACF  & \cite{rei+13} 	& \cite{nie+13} \\
      \hline
      \hline
      Total Number   &		 34,030	&	24,124			&		12,515	\\
      \hline
      \hline
      AND AutoACF  &   --  		&	20,009		& 	10,381		\\
       \hline
      AND \cite{rei+13} &	20,009	&    --   		 &	9,292		\\
       \hline
      AND \cite{nie+13}  &	10,381	&	9,292		&	--		\\
       \hline
 \end{tabular}
 \label{tab:comp_tab}
\end{table*}

\begin{figure*}[h]
  \centering
  \includegraphics[width=0.9\linewidth]{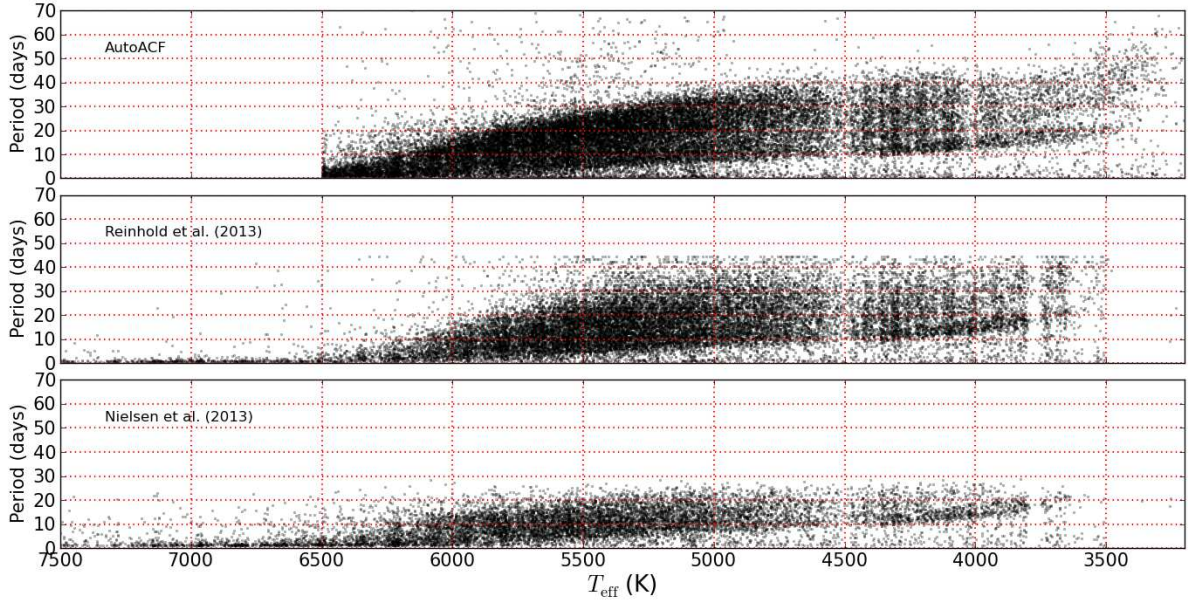}
  \caption{The $T_{\rm eff}$-period distributions of AutoACF (top), \cite{rei+13} (middle) and \cite{nie+13} (bottom). A discussion of the comparisons between the different methods can be found in Section~\ref{sec:comp_sub}.}
  \label{fig:tpc1}
\end{figure*}

\begin{figure*}[h]
  \centering
  \includegraphics[width=0.7\linewidth]{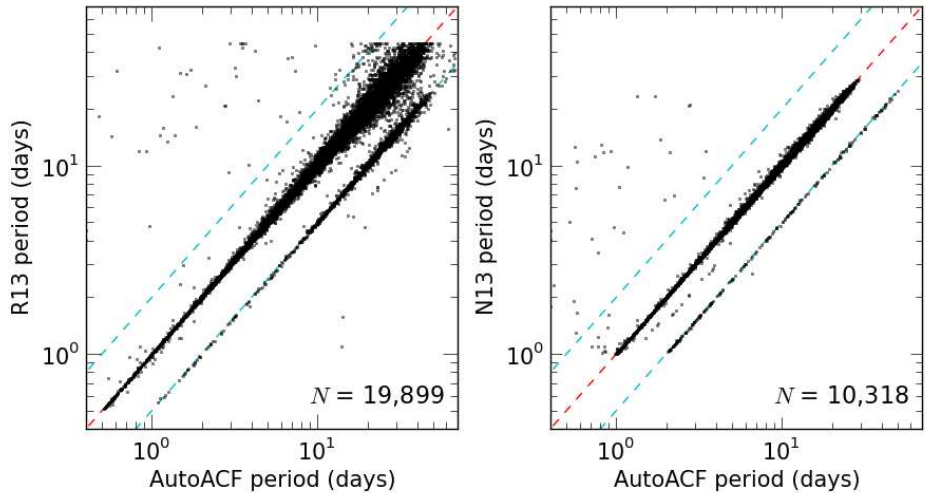}
  \caption{Direct comparisons between the periods measured in this work using AutoACF, and those detected by other groups. The panels show AutoACF compared to \cite{rei+13} (R13, left) and \cite{nie+13} (N13, right), with the number of stars in each group denoted as $N$. The red dashed line marks 1:1 ratio and the blue dashed lines mark 1:2 and 2:1.}
  \label{fig:tpc2}
\end{figure*}

During the course of this work, two rotation period samples from the \kepler\ data were published by \cite{rei+13} and \cite{nie+13}. In this section we summarize their methods and compare their results to that of AutoACF.

\subsection{Method of \cite{rei+13}}
\cite{rei+13} used Q3 PDC-MAP data and selected stars with $\log g > 3.5$ and variability amplitude $R_{\rm varB} \geq 0.003$, where $R_{\rm varB}$ \citep{bas+10,bas+11} is the $5^{\rm th}$ and $95^{\rm th}$ percentile of normalized flux, with 4\,hr boxcar smoothing of the LC. They applied the Generalized Lomb-Scargle periodogram \citep{zec+09} to LCs binned to 2\,hr sampling, in order to reduce computation time, and accepted periods in the range 12\,hrs--45\,days. Periods close to the orbital periods of known binaries were also excluded.

To minimize the aliasing caused by active regions on opposite hemispheres of the star (discussed in more detail in \cite{mcq+13a}, they compared the first two periods detected by the global sine fit, and if the difference between the longer period and twice the shorter period was less than \%5, then the longer period was selected. 

An automatic filter was applied based on the number of zero crossings per period in the smoothed LC, under the assumption that real stellar rotation signals should have a low number of zero crossings, whereas pulsations and irregular variations will have a higher number. In total, 24,124 rotation periods were accepted by these filters. 

\subsection{Method of \cite{nie+13}}

\cite{nie+13} used Q2--Q9 of the PDC-MAP and applied a Lomb-Scargle periodogram \citep{fra+95}. Known EBs and planet candidates were excluded from their sample, in addition to stars with $\log g < 3.5$. They considered a periodogram peak to be valid if it is greater than 4 times the white noise estimated from the root mean square (RMS) of the time series, and accepted periods in the range 1--30\,days.

Their selection routine required periods to be stable across multiple quarters, and the rotation period was then taken to be the median of the stable periods. They derived rotation periods for a total of 12,151 stars.

\subsection{Comparison with AutoACF}
\label{sec:comp_sub}

A comparison between the number of rotation period detections by each method is displayed in Table~\ref{tab:comp_tab}, showing that AutoACF detects significantly more periods than the alternative methods. In some cases, AutoACF did not detect a period and the other methods did. This occurred in 4,115 cases when compared to \cite{rei+13} and 1,770 compared to \cite{nie+13}. However, they had different initial sample selection criteria and when we compare only the stars to which AutoACF was applied, these numbers fall to 1,503 and 259 respectively. The cases missed by AutoACF with period detections from one of the other methods are predominantly those which failed to show coherent periods in multiple segments of the LC, and the remainder have low LPHs and fall just below our $w_{\rm thres}$. 
 
To compare our results to those of \cite{rei+13} and \cite{nie+13} we plotted period versus $T_{\rm eff}$ for each, which is shown in Figure~\ref{fig:tpc1}. A similar structure is displayed in all three cases, with AutoACF extending to longer periods. All methods find a sequence of increasing period with decreasing temperature, with a small fraction of fast rotators at all temperatures. 

The bimodality is clear in both the AutoACF and \cite{rei+13} plots, and would likely have been found also by \cite{nie+13} if longer periods were studied. Slight differences in the distribution below $\sim 4000$\,K, most notably the vertical artifacts and lower $T_{\rm eff}$ range, occur as a result of our use of the \cite{dre+13} parameters in place of the KIC values, where available.

To directly compare our periods with those detected by \cite{rei+13} and \cite{nie+13}, for the stars where both methods report periods, we plotted their periods versus ours in Figure~\ref{fig:tpc2}. We find that in 13,035 and 9,757 cases respectively, the AutoACF period is within $0.05\,P$ of that found by the alternative method, where $P$ is the AutoACF period. However, it is immediately clear that in some cases, both alternative methods detect half the period reported by AutoACF.  The periods of \cite{rei+13} are within $0.05\,(P/2)$ of half the AutoACF period in 1,304 cases (6.5\% of their periods), and the same is true for for the periods of  \cite{nie+13} in 202 cases (1.9\% of their periods). We manually checked that in these cases, the AutoACF reflects the visually confirmed period and the alternative methods have detected incorrect periods. We note that for periods below 20\,days, there is no evidence for AutoACF detecting half the period found by the other methods, however, for a small set of longer-period stars, \cite{rei+13} find twice the AutoACF period.

For the alternative methods there are $\sim 85$ cases where AutoACF detects a short period and \cite{rei+13} or \cite{nie+13} detect a considerably longer period. These cases are predominantly caused by multiple periods in the LC, where AutoACF has selected the short period and the alternative methods have selected the longer period. The rest appear to be false detections by the alternative methods, most often arising due to the short baseline used by \cite{rei+13}.

\section{Discussion}
\label{sec:disc}

The present catalog, of more than 30,000 rotation periods and photometric amplitudes, was derived using a systematic automated search from a well defined sample of more than 133,000 main-sequence stars observed by \kepler, all of which have stellar temperature estimates. This sample can help us study the mass-age-period relation, and better understand its implication on the amplitude of photometric variability.  

One clear feature of the sample is the fraction of rotation period detections as a function of temperature, which changes dramatically, from $\sim 80$\% for the cool stars to $\sim 20$\% for the hot stars of the sample. It seems that the photometric amplitude of the stellar rotation signature for the majority of the hot stars in the sample is smaller than our detection limit. This is supported by Figure~\ref{fig:teff_amp}, which shows that the detected amplitudes for the hot stars have a range that goes down to $\sim 300$\,ppm, below which we do not detect periodicity. For the cool stars, on the other hand,  the lower limit of the amplitudes is $\sim 2000$\,ppm, well above our detection threshold. We might suggest that the incompleteness of our detections for the cool stars is therefore related to the orientation angle of the stellar axis of rotation relative to our line of sight, which prevents us from detecting the rotation of all cool stars in the sample. 

Obviously, the photometric amplitude also carries information on the stellar surface inhomogeneity, presumably associated with the stellar magnetic field. Small amplitudes of variability can either result from smaller star spots or from a more even coverage of the stellar surface by spots. Therefore, our sample suggests that the typical inhomogeneity decreases as a function of stellar surface temperature. Further, Figure~\ref{fig:amp_per} shows that even for a given stellar temperature, the photometric amplitude is a decreasing function of the rotation period. This implies that the stellar inhomogeneity, probably together with the magnetic field, is decreasing as the star is aging, if we adopt the stellar rotation period as an age indicator. 

The catalog shows that the typical rotation period increases with decreasing stellar mass and temperature. A comparison to empirical gyrochronology models \citep{bar07, mam+08, mei+09} shows the upper envelope of points is broadly consistent with an age of 4.5\,Gyrs, but that the models appear to under-predict the stellar ages, with a significant proportion falling below the 1\,Gyr rotational isochrone. The position of the Sun close to the upper envelope is interesting, since this implies a lack of stars older than the Sun. More slowly rotating stars should still appear on the main sequence and would not have been removed by our exclusion of evolved stars.

The `dip' feature in the upper envelope around $\sim~0.5\,M_{\odot}$ ($\sim 4000$\,K), mentioned in \cite{mcq+13a} is still visible in the full sample. This mass corresponds to the transition point in young clusters, where lower mass stars spin up considerably more than stars above this mass \citep{hen+12}. This feature is not accounted for by any current gyrochronology models.

Thorough testing of AutoACF with synthetic LCs containing a range of periods and amplitudes is still required to understand the effect of observational biases on the upper envelope, since low-amplitude, long-period signals are most strongly affected by systematics and quarter joining.

A series of period histograms across varying $T_{\rm eff}$ bins, shown in Figure~\ref{fig:hists}, displays some of the finer details of the period distribution, including a significant bimodality in the low-mass stars, which becomes less apparent towards higher masses. This bimodality, forming two sequences in the mass- or $T_{\rm eff}$-period distribution, cannot be explained by any observational or measurement biases, and thus requires a physical explanation. One hypotheses presented in \cite{mcq+13a} is that this is a possible hint of two epochs of star formation in the solar neighborhood. This effect may become less apparent towards higher masses, since these brighter stars are seen to greater distances, hence probing a larger region of space. 

An alternative explanation, also presented in \cite{mcq+13a}, is that the dearth of stars between the two sequences represents a mass-dependent, rapid transition phase in the stellar spin-down law, where few stars are observed. In this scenario the star formation rate is unimodal but the spin-down occurs as two slow stages, with a rapid stage in between, producing a bimodal distribution of periods.  Both these hypotheses require further follow-up observations and detailed modeling \cite[e.g.,][]{gal+13, eps+14}, which will be aided by the larger sample size and wider $T_{\rm eff}$ coverage of the new data.

\acknowledgments
\noindent {\it Acknowledgments:} 
The authors wish to thank Jerome Bouvier, Arieh Konigl, Jennifer Van Saders and Jason Steffen for insightful discussions on this work. The research leading to these results has received funding from the European Research Council under the EU's Seventh Framework Programme (FP7/(2007-2013)/ERC Grant Agreement No.~291352). T.M. also acknowledges support from the Israel Science Foundation (grant No.~1423/11) and the Israeli Centers of Research Excellence (I-CORE, grant No.~1829/12). S.A. acknowledges support from STFC Consolidated grant ref. ST/K00106X/1 and Leverhulme Research Project grant RPG-2012-661. All of the data presented in this paper were obtained from the Mikulski Archive for Space Telescopes (MAST). STScI is operated by the Association of Universities for Research in Astronomy, Inc., under NASA contract NAS5-26555. Support for MAST for non-HST data is provided by the NASA Office of Space Science via grant NNX09AF08G and by other grants and contracts. 


\begin{thebibliography}{}
\expandafter\ifx\csname natexlab\endcsname\relax\def\natexlab#1{#1}\fi

\bibitem[{Affer {et~al.}(2012)Affer, Micela, Favata, \& Flaccomio}]{aff+12}
Affer, L., Micela, G., Favata, F., \& Flaccomio, E. 2012, MNRAS, 424, 11

\bibitem[{{Akeson} {et~al.}(2013){Akeson}, {Chen}, {Ciardi}, {Crane}, {Good},
  {Harbut}, {Jackson}, {Kane}, {Laity}, {Leifer}, {Lynn}, {McElroy}, {Papin},
  {Plavchan}, {Ram{\'{\i}}rez}, {Rey}, {von Braun}, {Wittman}, {Abajian},
  {Ali}, {Beichman}, {Beekley}, {Berriman}, {Berukoff}, {Bryden}, {Chan},
  {Groom}, {Lau}, {Payne}, {Regelson}, {Saucedo}, {Schmitz}, {Stauffer},
  {Wyatt}, \& {Zhang}}]{ake+13}
{Akeson}, R.~L., {Chen}, X., {Ciardi}, D., {et~al.} 2013, \pasp, 125, 989

\bibitem[{{Baliunas} {et~al.}(1996){Baliunas}, {Sokoloff}, \& {Soon}}]{bal+96}
{Baliunas}, S., {Sokoloff}, D., \& {Soon}, W. 1996, ApJL, 457, L99

\bibitem[{{Baraffe} {et~al.}(1998){Baraffe}, {Chabrier}, {Allard}, \&
  {Hauschildt}}]{bar+98}
{Baraffe}, I., {Chabrier}, G., {Allard}, F., \& {Hauschildt}, P.~H. 1998, A\&A,
  337, 403

\bibitem[{Barnes(2003)}]{bar03}
Barnes, S.~A. 2003, ApJ, 586, 464

\bibitem[{Barnes(2007)}]{bar07}
---. 2007, ApJ, 669, 1167

\bibitem[{Barnes(2010)}]{bar10}
---. 2010, ApJ, 722, 222

\bibitem[{{Basri} {et~al.}(2010){Basri}, {Walkowicz}, {Batalha}, {Gilliland},
  {Jenkins}, {Borucki}, {Koch}, {Caldwell}, {Dupree}, {Latham}, {Meibom},
  {Howell}, \& {Brown}}]{bas+10}
{Basri}, G., {Walkowicz}, L.~M., {Batalha}, N., {et~al.} 2010, ApJL, 713, L155

\bibitem[{{Basri} {et~al.}(2011){Basri}, {Walkowicz}, {Batalha}, {Gilliland},
  {Jenkins}, {Borucki}, {Koch}, {Caldwell}, {Dupree}, {Latham}, {Marcy},
  {Meibom}, \& {Brown}}]{bas+11}
---. 2011, AJ, 141, 20

\bibitem[{Borucki {et~al.}(2010)Borucki, Koch, Basri, Batalha, Brown, Caldwell,
  Caldwell, Christensen-Dalsgaard, Cochran, DeVore, Dunham, Dupree, Gautier,
  Geary, Gilliland, Gould, Howell, Jenkins, Kondo, Latham, Marcy, Meibom,
  Kjeldsen, Lissauer, Monet, Morrison, Sasselov, Tarter, Boss, Brownlee, Owen,
  Buzasi, Charbonneau, Doyle, Fortney, Ford, Holman, Seager, Steffen, Welsh,
  Rowe, Anderson, Buchhave, Ciardi, Walkowicz, Sherry, Horch, Isaacson,
  Everett, Fischer, Torres, Johnson, Endl, MacQueen, Bryson, Dotson, Haas,
  Kolodziejczak, Van~Cleve, Chandrasekaran, Twicken, Quintana, Clarke, Allen,
  Li, Wu, Tenenbaum, Verner, Bruhweiler, Barnes, \& Prsa}]{bor+10}
Borucki, W.~J., Koch, D., Basri, G., {et~al.} 2010, Science, 327, 977

\bibitem[{{Bouvier}(2013)}]{bou+13}
{Bouvier}, J. 2013, in EAS Publications Series, Vol.~62, EAS Publications
  Series, 143--168

\bibitem[{Bouvier {et~al.}(1997)Bouvier, Forestini, \& Allain}]{bou+97}
Bouvier, J., Forestini, M., \& Allain, S. 1997, A\&A, 326, 1023

\bibitem[{{Ciardi} {et~al.}(2011){Ciardi}, von Braun, Bryden, van Eyken,
  Howell, Kane, Plavchan, Ram{\'\i}rez, \& Stauffer}]{cia+11}
{Ciardi}, D.~R., von Braun, K., Bryden, G., {et~al.} 2011, AJ, 141, 108

\bibitem[{Collier~Cameron {et~al.}(2009)Collier~Cameron, Davidson, Hebb,
  Skinner, Anderson, Christian, Clarkson, Enoch, Irwin, Joshi, Haswell,
  Hellier, Horne, Kane, Lister, Maxted, Norton, Parley, Pollacco, Ryans,
  Scholz, Skillen, Smalley, Street, West, Wilson, \& Wheatley}]{col+09}
Collier~Cameron, A., Davidson, V.~A., Hebb, L., {et~al.} 2009, MNRAS, 400, 451

\bibitem[{{Dressing} \& {Charbonneau}(2013)}]{dre+13}
{Dressing}, C.~D., \& {Charbonneau}, D. 2013, \apj, 767, 95

\bibitem[{{Epstein} \& {Pinsonneault}(2014)}]{eps+14}
{Epstein}, C.~R., \& {Pinsonneault}, M.~H. 2014, \apj, 780, 159

\bibitem[{{Frandsen} {et~al.}(1995){Frandsen}, {Jones}, {Kjeldsen}, {Viskum},
  {Hjorth}, {Andersen}, \& {Thomsen}}]{fra+95}
{Frandsen}, S., {Jones}, A., {Kjeldsen}, H., {et~al.} 1995, \aap, 301, 123

\bibitem[{{Gallet} \& {Bouvier}(2013)}]{gal+13}
{Gallet}, F., \& {Bouvier}, J. 2013, \aap, 556, A36

\bibitem[{{Gizon}(2002)}]{giz02}
{Gizon}, L. 2002, Astronomische Nachrichten, 323, 251

\bibitem[{{Gizon} \& {Solanki}(2004)}]{giz+04}
{Gizon}, L., \& {Solanki}, S.~K. 2004, \solphys, 220, 169

\bibitem[{{Goulding} {et~al.}(2012){Goulding}, {Barnes}, {Pinfield},
  {Kov{\'a}cs}, {Birkby}, {Hodgkin}, {Catal{\'a}n}, {Sip{\H o}cz}, {Jones},
  {Del Burgo}, {Jeffers}, {Nefs}, {G{\'a}lvez-Ortiz}, \& {Martin}}]{gou+12}
{Goulding}, N.~T., {Barnes}, J.~R., {Pinfield}, D.~J., {et~al.} 2012, MNRAS,
  427, 3358

\bibitem[{{Harrison} {et~al.}(2012){Harrison}, {Coughlin}, {Ule}, \&
  {L{\'o}pez-Morales}}]{har+12}
{Harrison}, T.~E., {Coughlin}, J.~L., {Ule}, N.~M., \& {L{\'o}pez-Morales}, M.
  2012, AJ, 143, 4

\bibitem[{{Hartigan} \& {Hartigan}(1985)}]{hart85}
{Hartigan}, J.~A., \& {Hartigan}, P.~M. 1985, The Annals of Statistics, 13, 70

\bibitem[{{Hartman} {et~al.}(2011){Hartman}, {Bakos}, {Noyes}, {Sip{\H o}cz},
  {Kov{\'a}cs}, {Mazeh}, {Shporer}, \& {P{\'a}l}}]{har+11}
{Hartman}, J.~D., {Bakos}, G.~{\'A}., {Noyes}, R.~W., {et~al.} 2011, \aj, 141,
  166

\bibitem[{{Henderson} \& {Stassun}(2012)}]{hen+12}
{Henderson}, C.~B., \& {Stassun}, K.~G. 2012, \apj, 747, 51

\bibitem[{{Huber} {et~al.}(2014){Huber}, {Silva Aguirre}, {Matthews},
  {Pinsonneault}, {Gaidos}, {Garc{\'{\i}}a}, {Hekker}, {Mathur}, {Mosser},
  {Torres}, {Bastien}, {Basu}, {Bedding}, {Chaplin}, {Demory}, {Fleming},
  {Guo}, {Mann}, {Rowe}, {Serenelli}, {Smith}, \& {Stello}}]{hub+13}
{Huber}, D., {Silva Aguirre}, V., {Matthews}, J.~M., {et~al.} 2014, \apjs, 211,
  2

\bibitem[{{Irwin} {et~al.}(2009){Irwin}, {Aigrain}, {Bouvier}, {Hebb},
  {Hodgkin}, {Irwin}, \& {Moraux}}]{irw+09a}
{Irwin}, J., {Aigrain}, S., {Bouvier}, J., {et~al.} 2009, \mnras, 392, 1456

\bibitem[{Irwin {et~al.}(2011)Irwin, Berta, Burke, Charbonneau, Nutzman, West,
  \& Falco}]{irw+11}
Irwin, J., Berta, Z.~K., Burke, C.~J., {et~al.} 2011, ApJ, 727, 56

\bibitem[{{Irwin} \& {Bouvier}(2009)}]{irw+09}
{Irwin}, J., \& {Bouvier}, J. 2009, in IAU Symp., Vol. 258, The Ages of Stars,
  363--374

\bibitem[{{Kaler}(1989)}]{kal89}
{Kaler}, J.~B. 1989, {Stars and their spectra. an introduction to spectral
  sequence} ({Cambridge: Cambridge University Press})

\bibitem[{Kawaler(1988)}]{kaw88}
Kawaler, S.~D. 1988, ApJ, 333, 236

\bibitem[{Kawaler(1989)}]{kaw89}
---. 1989, ApJ, 343, L65

\bibitem[{{Kenyon} \& {Hartmann}(1995)}]{ken+95}
{Kenyon}, S.~J., \& {Hartmann}, L. 1995, \apjs, 101, 117

\bibitem[{{Kiraga} \& {Stepien}(2007)}]{kist07}
{Kiraga}, M., \& {Stepien}, K. 2007, ACTAA, 57, 149

\bibitem[{{Koch} {et~al.}(2010){Koch}, {Borucki}, {Basri}, {Batalha}, {Brown},
  {Caldwell}, {Christensen-Dalsgaard}, {Cochran}, {DeVore}, {Dunham},
  {Gautier}, {Geary}, {Gilliland}, {Gould}, {Jenkins}, {Kondo}, {Latham},
  {Lissauer}, {Marcy}, {Monet}, {Sasselov}, {Boss}, {Brownlee}, {Caldwell},
  {Dupree}, {Howell}, {Kjeldsen}, {Meibom}, {Morrison}, {Owen}, {Reitsema},
  {Tarter}, {Bryson}, {Dotson}, {Gazis}, {Haas}, {Kolodziejczak}, {Rowe}, {Van
  Cleve}, {Allen}, {Chandrasekaran}, {Clarke}, {Li}, {Quintana}, {Tenenbaum},
  {Twicken}, \& {Wu}}]{koc+10}
{Koch}, D.~G., {Borucki}, W.~J., {Basri}, G., {et~al.} 2010, \apjl, 713, L79

\bibitem[{{Mamajek}(2011)}]{mam11}
{Mamajek}, E. 2011, {A Modern Mean Stellar Color and Effective Temperatures
  (Teff) Sequence for O9V-Y0V Dwarf Stars}, Online: accessed 24-Feb-2013

\bibitem[{{Mamajek} \& {Hillenbrand}(2008)}]{mam+08}
{Mamajek}, E.~E., \& {Hillenbrand}, L.~A. 2008, \apj, 687, 1264

\bibitem[{{Matt} {et~al.}(2012){Matt}, {MacGregor}, {Pinsonneault}, \&
  {Greene}}]{mat+12}
{Matt}, S.~P., {MacGregor}, K.~B., {Pinsonneault}, M.~H., \& {Greene}, T.~P.
  2012, \apjl, 754, L26

\bibitem[{{McQuillan} {et~al.}(2012){McQuillan}, {Aigrain}, \&
  {Roberts}}]{mcq+12}
{McQuillan}, A., {Aigrain}, S., \& {Roberts}, S. 2012, \aap, 539, A137

\bibitem[{{McQuillan} {et~al.}(2013{\natexlab{a}}){McQuillan}, {Aigrain}, \&
  {Mazeh}}]{mcq+13a}
{McQuillan}, A., {Aigrain}, S., \& {Mazeh}, T. 2013{\natexlab{a}}, \mnras, 432,
  1203

\bibitem[{{McQuillan} {et~al.}(2013{\natexlab{b}}){McQuillan}, {Mazeh}, \&
  {Aigrain}}]{mcq+13b}
{McQuillan}, A., {Mazeh}, T., \& {Aigrain}, S. 2013{\natexlab{b}}, \apjl, 775,
  L11

\bibitem[{Meibom {et~al.}(2009)Meibom, Mathieu, \& Stassun}]{mei+09}
Meibom, S., Mathieu, R.~D., \& Stassun, K.~G. 2009, ApJ, 695, 679

\bibitem[{{Meibom} {et~al.}(2011){Meibom}, {Mathieu}, {Stassun}, {Liebesny}, \&
  {Saar}}]{mei+11}
{Meibom}, S., {Mathieu}, R.~D., {Stassun}, K.~G., {Liebesny}, P., \& {Saar},
  S.~H. 2011, \apj, 733, 115

\bibitem[{{Moraux} {et~al.}(2013){Moraux}, {Artemenko}, {Bouvier}, {Irwin},
  {Ibrahimov}, {Magakian}, {Grankin}, {Nikogossian}, {Cardoso}, {Hodgkin},
  {Aigrain}, \& {Movsessian}}]{mor+13}
{Moraux}, E., {Artemenko}, S., {Bouvier}, J., {et~al.} 2013, \aap, 560, A13

\bibitem[{{Nielsen} {et~al.}(2013){Nielsen}, {Gizon}, {Schunker}, \&
  {Karoff}}]{nie+13}
{Nielsen}, M.~B., {Gizon}, L., {Schunker}, H., \& {Karoff}, C. 2013, \aap, 557,
  L10

\bibitem[{{Pizzolato} {et~al.}(2003){Pizzolato}, {Maggio}, {Micela},
  {Sciortino}, \& {Ventura}}]{piz+03}
{Pizzolato}, N., {Maggio}, A., {Micela}, G., {Sciortino}, S., \& {Ventura}, P.
  2003, \aap, 397, 147

\bibitem[{{Pr{\v s}a} {et~al.}(2011){Pr{\v s}a}, {Batalha}, {Slawson}, {Doyle},
  {Welsh}, {Orosz}, {Seager}, {Rucker}, {Mjaseth}, {Engle}, {Conroy},
  {Jenkins}, {Caldwell}, {Koch}, \& {Borucki}}]{prs+11}
{Pr{\v s}a}, A., {Batalha}, N., {Slawson}, R.~W., {et~al.} 2011, \aj, 141, 83

\bibitem[{{Reiners} \& {Mohanty}(2012)}]{rei+12}
{Reiners}, A., \& {Mohanty}, S. 2012, \apj, 746, 43

\bibitem[{{Reinhold} {et~al.}(2013){Reinhold}, {Reiners}, \& {Basri}}]{rei+13}
{Reinhold}, T., {Reiners}, A., \& {Basri}, G. 2013, \aap, 560, A4

\bibitem[{{Sekiguchi} \& {Fukugita}(2000)}]{sek+00}
{Sekiguchi}, M., \& {Fukugita}, M. 2000, \aj, 120, 1072

\bibitem[{{Slawson} {et~al.}(2011){Slawson}, {Pr{\v s}a}, {Welsh}, {Orosz},
  {Rucker}, {Batalha}, {Doyle}, {Engle}, {Conroy}, {Coughlin}, {Gregg},
  {Fetherolf}, {Short}, {Windmiller}, {Fabrycky}, {Howell}, {Jenkins}, {Uddin},
  {Mullally}, {Seader}, {Thompson}, {Sanderfer}, {Borucki}, \& {Koch}}]{sla+11}
{Slawson}, R.~W., {Pr{\v s}a}, A., {Welsh}, W.~F., {et~al.} 2011, \aj, 142, 160

\bibitem[{{Smith} {et~al.}(2012){Smith}, {Stumpe}, {Van Cleve}, {Jenkins},
  {Barclay}, {Fanelli}, {Girouard}, {Kolodziejczak}, {McCauliff}, {Morris}, \&
  {Twicken}}]{smi+12}
{Smith}, J.~C., {Stumpe}, M.~C., {Van Cleve}, J.~E., {et~al.} 2012, \pasp, 124,
  1000

\bibitem[{{Stumpe} {et~al.}(2012){Stumpe}, {Smith}, {Van Cleve}, {Twicken},
  {Barclay}, {Fanelli}, {Girouard}, {Jenkins}, {Kolodziejczak}, {McCauliff}, \&
  {Morris}}]{stu+12}
{Stumpe}, M.~C., {Smith}, J.~C., {Van Cleve}, J.~E., {et~al.} 2012, \pasp, 124,
  985

\bibitem[{{van Saders} \& {Pinsonneault}(2013)}]{vs+13}
{van Saders}, J.~L., \& {Pinsonneault}, M.~H. 2013, \apj, 776, 67

\bibitem[{{Walkowicz} \& {Basri}(2013)}]{wal+13}
{Walkowicz}, L.~M., \& {Basri}, G.~S. 2013, \mnras, 436, 1883

\bibitem[{{Zechmeister} \& K{\"u}rster(2009)}]{zec+09}
{Zechmeister}, M., \& K{\"u}rster, M. 2009, A\&A, 496, 577

\end{thebibliography}
\bibliographystyle{apj}

\appendix{}
\section{Section A: Automation of the ACF}
\label{sec:auto}
Our previous work using the ACF method has relied upon visual verification of the period detection of each LC \citep{mcq+13a, mcq+13b}. The large number of stars in the full \kepler\ sample makes this approach unfeasible, and therefore we opted to design an automated period verification procedure. In this appendix we describe the design and testing of the automated ACF (AutoACF) method.

\subsection{Visually Examined Training Set}
\label{sec:vis_ex}

A large subset of \kepler\ targets were visually examined to form a training set for the automation procedure. During this process, each visually examined target was assigned a classification:
\begin{enumerate}
	\item{Periodic: Clear periodicity in LC, with the selected ACF peak matching the visually identified period.}
	\item{Wrong Peak: Clear periodicity in LC but incorrect peak selected in ACF (usually half or twice the rotation period identified by eye).}
	\item{Probable Periodic: Some indication in the LC and ACF that the star is periodic, but not clear enough to include in the `Periodic' training set.}
	\item{Likely Pulsation: The LC and ACF indicate that the periodicity detected is more likely to arise from effects other than rotation. See Section 3.5 of \cite{mcq+13a} for a more detailed discussion.}
	\item{Non Periodic: The LC and ACF show no signs of periodicity, or the ACF peak clearly results from instrumental systematics.}
\end{enumerate}

\begin{figure}[h]
  \centering
  \includegraphics[width=\linewidth]{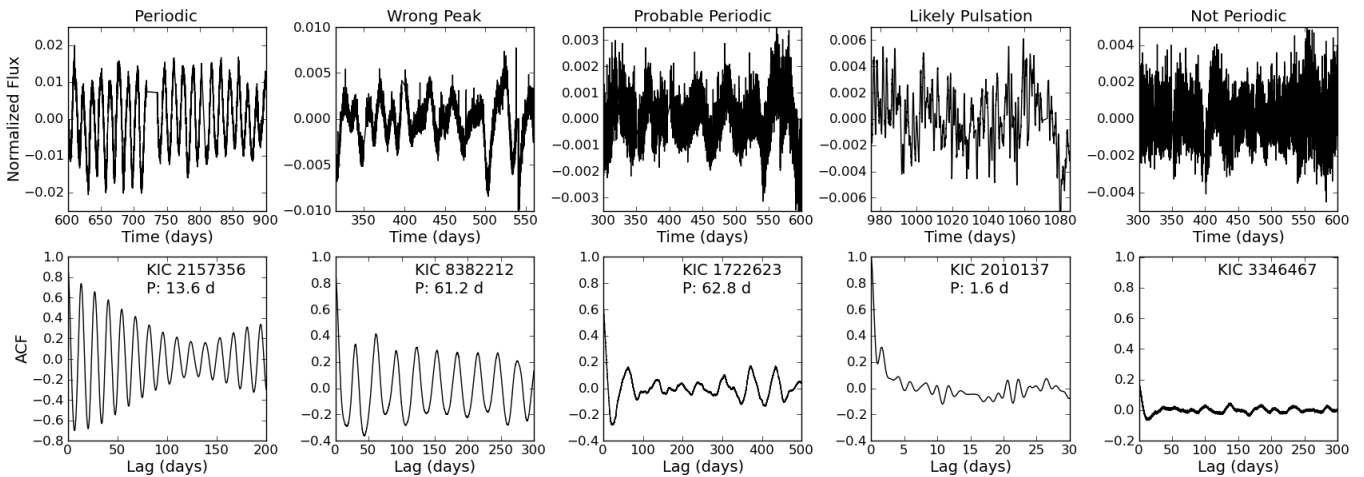}
  \caption{Example LCs (top) and ACFs (bottom) of the different classifications listed in Section~\ref{sec:vis_ex}, with the classification type noted above each column.}
  \label{fig:train_ex}
\end{figure}

Figure~\ref{fig:train_ex} shows examples of these 5 classes: The first column shows a clear case of periodicity; the second column shows an example where the rotation period can be identified by eye as $\sim 30$ days, but due to residual systematics in the LC, the second ACF peak is higher and therefore erroneously selected to represent the period; the third column shows a case which is likely to be periodic based on the repeated features in the LC and the peaks in the ACF, but is not clear enough to make it into the `Period' set; the fourth column shows rapid and irregular variability which gives rise to irregular peaks in the ACF, which we consider unlikely to arise from rotational modulation; and the final column shows a case with a noise-dominated LC and no obvious peaks in the ACF.

We divided the sample into 500\,K $T_{\rm eff}$ bins, and visually examined the LCs and ACFs in a random order until 1000 verified periods (categories 1 and 2) were classified in each bin. The results of the above identification process are shown in Table~\ref{tab:trainset}. 

At this stage we tested a variety of statistics to distinguish automatically between the visually confirmed detection and the false positives. This included using the height, width and number of the detected ACF peaks, the uncertainty on the period measurements, and the stability of the period detection over different quarters. Ultimately, we converged on the two-stage selection method described below.

\begin{table}[h]
  \caption{Training set selection: Temperature bin, total number of targets, number visually examined targets,  the visual classifications described in the text, and the periodic fraction of the visually examined set.}
  \centering
  \begin{tabular}{c | c | c | c cccc |c} 
    \hline
      $T_{\rm eff}$ & Total & Visually Examined & Periodic & Wrong Peak & Probable Periodic  & Pulsation & Non Periodic & Periodic \%\\
    \hline  
     3500--4000   &   1657 &  1657   & 1342    & 26  & 181  &  47    &  61 &  81.0 \\  
     4000--4500   &   3875 &  1466   & 994  & 6   &  293       & 31     & 142 &  68.2 \\ 
     4500--5000   &   7761 &  2105   & 987  & 13  & 417       & 75     & 613  &  47.5 \\   
     5000--5500   &   23767 & 3337     & 988  & 12  & 737   & 64     & 1536 &  30.0 \\  
     5500--6000   &   37839 &  4580   & 985  & 15  & 797     & 74    & 2709 &   21.8\\  
     6000--6500   &   14565 &  3644    & 978  &  22  & 589  & 82     & 1973  &  27.4\\  
     \hline
     Total   		  &   89464 & 16789 & 6274 & 94 & 3014 & 373    & 7034 & 37.4    \\
    \hline
 \end{tabular}
 \label{tab:trainset}
\end{table}

\subsection{Selection Process 1 - Coherent Periods in Multiple LC Segments}
The separation of periodic and non-periodic ACF results was performed in two steps. A real astrophysical periodicity is more likely to be detected in multiple regions of the LC than a systematic effect or artifact would. Our first selection procedure was therefore to accept only ACF detections that are consistent in different parts of the LC. We adopted the criterion that the period detected in the full Q3--Q14 LC must be coherent with that found in 2 out of the 4 segments of the data, where a segment is defined as 3 consecutive \kepler\ quarters (Q3--Q5, Q6--Q8, Q9--Q11, Q12--Q14). The period detection in a segment is defined as coherent if it is within 10\% of the period (or half / twice the period) detected in the full Q3--Q14 LC. 

This method selected 63.8\% of the training sample, of which 63.9\% were classed as `Periodic'. This selection retained 99.3\% of the `Periodic' classifications from the training set.  The multiple segment criterion is not sufficient alone to distinguish between true and false period detections. Due to the rotation of the \kepler\ satellite through $90^{\circ}$ between quarters, the systematics can repeat every 4 quarters when targets appear in the same place on the CCD. This can lead to matching, but false period detections in multiple segments. It is also possible to have low-amplitude ACF peaks arising from noise, which are consistent by coincidence over multiple segments. We therefore required a second selection stage.

\subsection{Selection Process 2 - ACF Local Peak Height Weighting}

The local peak height (LPH) of the ACF is defined as the height of the selected peak with respect to the mean of the troughs on either side. Visually confirmed period detections typically have higher LPHs than those suspected to originate from noise or systematic effects. The LPH is also correlated with period, where shorter periods have higher LPHs. This is a result of the increased effect of instrumental systematics and LC processing on longer periods. 

To exploit the properties of LPH for distinguishing between periodic and non-periodic ACFs we assign a weight, $w$, between 0--1, based on LPH, period and $T_{\rm eff}$, where a higher weight indicates a greater probability of a true period detection. This weight comprises two parts, which are described below and shown in Figure~\ref{fig:w12}. A threshold, $w_{\rm thres}$ can then be applied to distinguish between periodic and false detections. $T_{\rm eff}$ is incorporated into the selection criteria because the visually confirmed period distribution varies with temperature, and hence the probability of a detection being real at a certain period and LPH is different for targets of different $T_{\rm eff}$.

\subsubsection{Weight Part 1: Magnitude of LPH}

The first part of the weight, $w_1$, is dependent on the magnitude of LPH, adjusted by a positive gradient to account for the fact that at longer periods, higher LPHs are found for false detections than at short periods. If the LC has systematic effects in two or three quarters, these can easily produce long-period ACF peaks with high LPHs, whereas short period effects would require many more coherent repetitions to produce an ACF peak with the same LPH, and hence have a higher likelihood of being real detections. The baseline, marked as the dashed line in each panel of Figure~\ref{fig:w12}, is described by the equation: $\mbox{LPH}_{\rm line} =  0.0027 P + 0.126$, where $P$ is period. This line was found by fitting a straight line to the period of the `Not Periodic' points as a function of LPH for the middle two temperature bins (4500--5500\,K) and adding 0.05 to the y-offset to approximate the upper edge of points rather than the central fit: 

The $w_1$ value for each LC is then calculated from 
\begin{equation}
w_{1,i} = 1.0 - \exp[- (\mbox{LPH}_{i} - \mbox{LPH}_{\rm line})],
\end{equation} 
where negative values of $(\mbox{LPH}_{i} - \mbox{LPH}_{\rm line})$ are set to 0, and $i = 0,1,2,..., N$ where $N$ is the number of LCs. This gives a  sharp increase in $w_1$ at low LPHs and then a more gradual increase at high LPH, which is required to separate true detection from the continuum of false detections. 

\subsubsection{Weight  Part 2: Position in $T_{\rm eff}$-Period-LPH Space}

\begin{figure}
  \centering
  \includegraphics[width=0.98\linewidth]{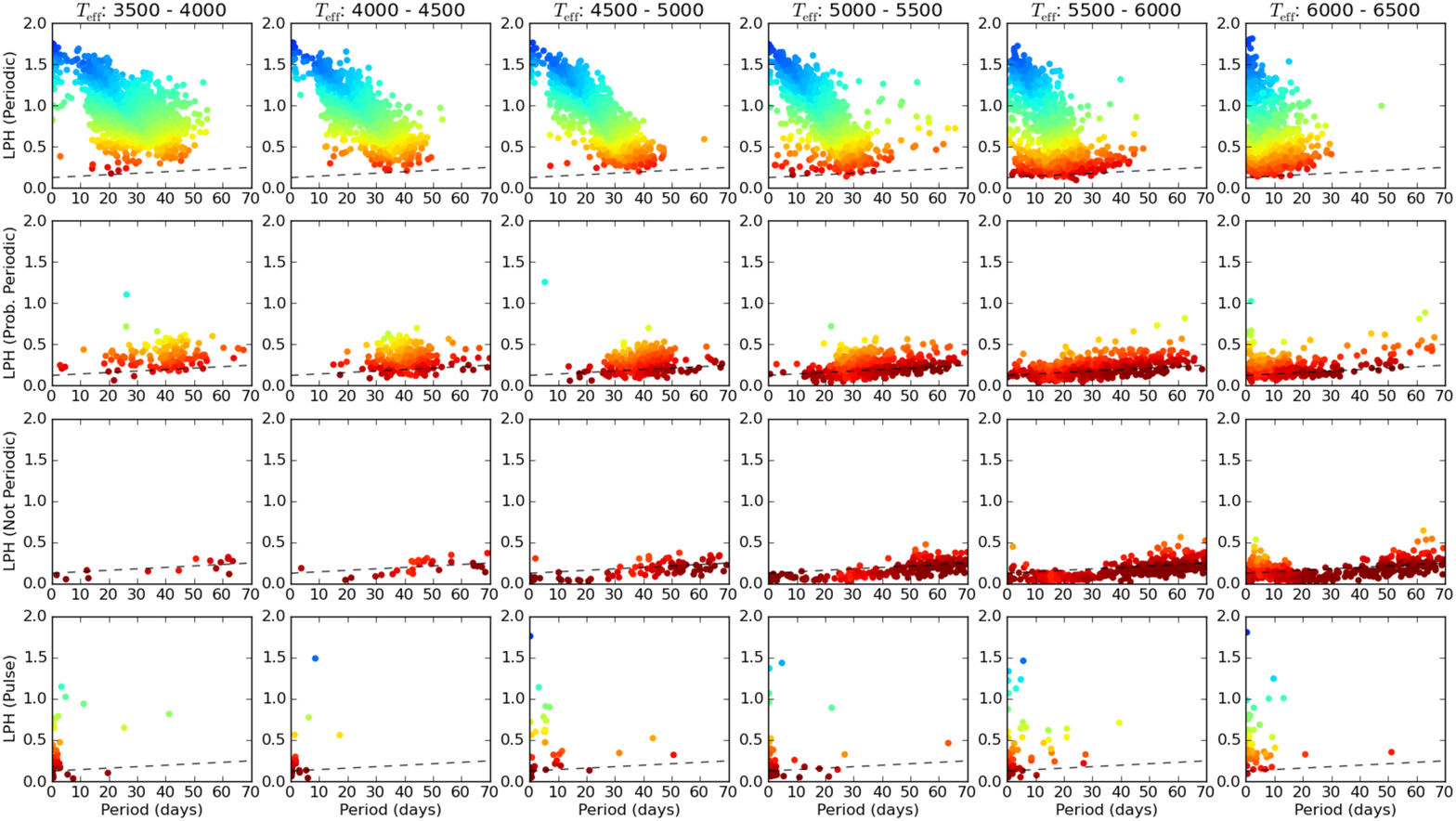}
  \includegraphics[width=0.7\linewidth]{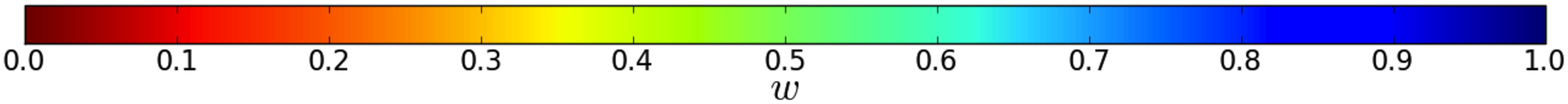}
  \caption{Each panel shows the period-LPH distribution, colored by total weight, $w$. The top row displays the `Periodic' class in each $T_{\rm eff}$ bin, the second row shows the `Probable Periodic' class, the third row is the `Non Periodic' class and the bottom row is the `Pulsation' class. The dashed line, described in the text, shows the baseline from which $w_1$ is measured.}
   \label{fig:w12}
\end{figure}

The second part of the weighting, $w_2$, is based on the location of the target in $T_{\rm eff}$-period-LPH space. Since `Periodic' detections typically lie along a sequence of decreasing LPH with increasing period (see Figure~\ref{fig:w12}), detections further from this band are more likely to be false. This is especially important at low LPHs, where a `Periodic' detection at the base of the periodic sequence can have the same LPH as a `Non Periodic' detection further from the base of the sequence. We therefore identified the position and width of this sequence as a function of $T_{\rm eff}$, and developed a criterion for selection based on distance from the expected location in $T_{\rm eff}$-period-LPH space. 

The position and width, $\alpha$,  of the sequence were found using a polynomial fit to the median, $5^{th}$ and $95^{th}$ percentiles of the period at intervals in LPH for each temperature bin. We used linear interpolation between the temperature bins to obtain an expected period, $P_{\rm ex}$, as a function of $T_{\rm eff}$ and LPH, and $\alpha$ as a function of $T_{\rm eff}$.

The distance from this expected period, for a given LPH measurement is converted to $w_2$ by:
\begin{equation}
w_{2, i} = \frac{\exp(- (d_{i} / \alpha_i)}{\mbox{MAX}(\exp(-(2d_{i} / \alpha_i)},
\end{equation}
where $d_i = |P_i - P_{i,{\rm ex}}|$. By combining $w_1$ and $w_2$, an overall weight $w$ is obtained, which has a range between 0--1, with higher values at higher LPHs, with the exception of close to the expected period, where the weight is boosted even for relatively low LPHs. The weights combined as $w = (w_1 + C w_2) / C$, where $C$ is a parameter of the automation routine which is adjusted to give the maximum separation between detection and false positives from the training set. The resulting distribution of $w$ is displayed in Figure~\ref{fig:w12}, which shows the training set results. To avoid excess false positives from pulsators and unfeasibly long periods, we only consider periods in the range 0.2--70 days. This range was chosen by visual examination of the LCs in the training set. 

The parameter $C$, which controls the ratio of the weight combination, and the threshold $w_{\rm thres}$ were selected to optimize the distinction between periodic and false positives. Our priority was primarily to minimize the number of false positives (`Non Periodic' class), while including the highest possible percentage of the `Periodic' class and a minimal number of the `Probable Periodic' class.  By running AutoACF on the training set, using a range of possible values for $C$ and $w_{\rm thres}$, we found that $C = 0.15$ and $w_{\rm thres} = 0.25$ produce the best match to our selection criteria. Tables~\ref{tab:main_res}~and~\ref{tab:rem_res} includes $w$ so that users of this data may select their own threshold to be more or less conservative. 

We considered the potential of the $T_{\rm eff}$-dependent component of the weighting to bias the result towards periods that fit the global trend, but concluded that the effect will not preferentially exclude periods outside the expected trend region, provided their LPH is sufficient indicate a real detection. The second weighting is designed to select additional real periods which are found to fall below a simple LPH threshold, as indicated by visual inspection. 

\subsection{Training Set Results}

Table~\ref{tab:train_res} shows the outcome of AutoACF applied to the training set, and demonstrates that the method is able to select 86.2\% of the visually identified `Periodic' class with only a 0.1\% false positive rate (`Non Periodic'). The fractions of each class in the selected group represent the predicted fractions obtained when applying AutoACF to the full \kepler\ sample.

\begin{table}[h]
  \caption{Training set results: Selected numbers per classification, using $w_{\rm crit} = 0.25$. The `Selected' row shows the number of each class determined as periodic by AutoACF. The percentages in brackets are percent of the selected sets consisting of each class, i.e., 94.5\% of the `Selected' set is the `Periodic' class.}
  \centering
  \begin{tabular}{c | ccccc | c} 
    \hline
    & Periodic &  Wrong Peak  & Probable Periodic & Pulsation  & Non Periodic & Total\\
    \hline
    Selected by AutoACF & 5411 (94.5\%)  & 32 (0.6\%)   & 191 (3.3\%)  & 82 (1.4\%) & 8 (0.1\%)  & 5724\\
    \hline
 \end{tabular}
 \label{tab:train_res}
\end{table}

\begin{figure}
  \centering
  \includegraphics[width=0.7\linewidth]{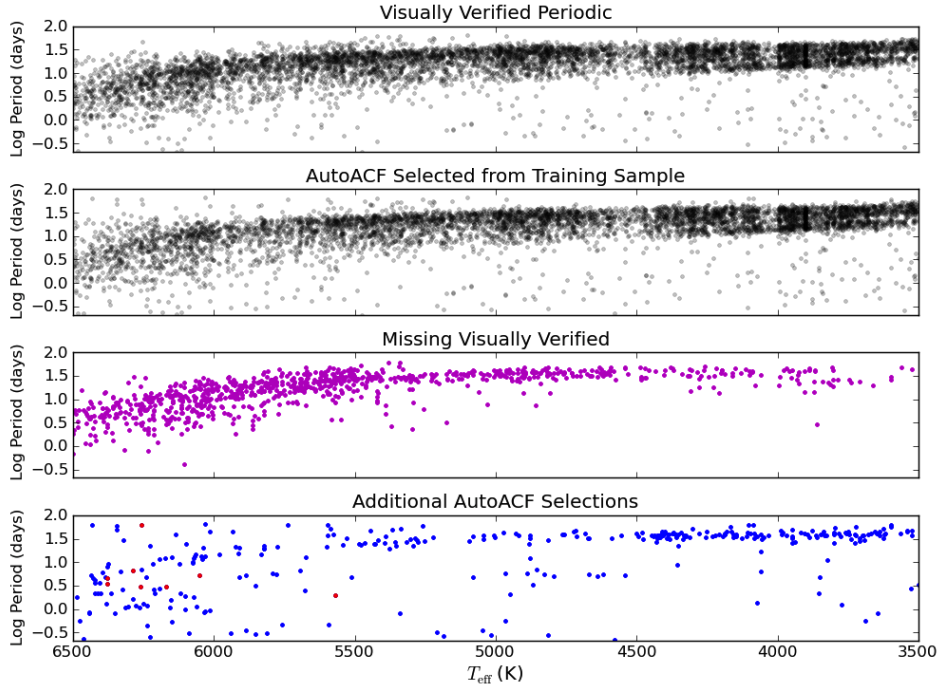}
  \caption{Training Set: Comparison between the visually examined `Periodic' class and the AutoACF selection of the visually examined sample. Top panel: Visually verified `Periodic' sample. Second panel: AutoACF selected sample. Third panel:  The `Periodic' class targets not in the AutoACF selection (magenta). Fourth panel: The `Probable Periodic' (blue) and `Non Periodic' (red) selected by AutoACF.}
  \label{fig:trainset}
\end{figure}

To explore the differences between the visually identified periodic set and the AutoACF selected members of the training set, we plotted comparisons in the $T_{\rm eff}$-period distribution (Figure~\ref{fig:trainset}). In each panel the sequence populated by real rotation period detections is clearly shown. The first and second panels directly compare the visually confirmed `Periodic' class and the AutoACF selected set, showing that the two differ very little from one another. The third and fourth panels display in color the detections that are different between the visual and AutoACF selection methods. The magenta points in the third panel are visually verified `Periodic' detections that are not selected by AutoACF.  The colored points in the fourth panel show AutoACF selected points that are not in the visually `Periodic' class. These primarily consist of `Probable Periodic' detections (cyan) with a small number of the `Non Periodic' class (blue). The additional `Probable Periodic' targets are typically long period detections, and their inclusion by AutoACF is expected, since longer period signals are more strongly affected by systematics and LC processing, which may cause the signal to be less clear to visual inspection. The differences between the visually inspected `Periodic' sample and the AutoACF sample do not significantly alter the shape of the period-$T_{\rm eff}$ distribution.

\section{Section B: Additional Detections}
\label{sec:add}

\begin{figure}
  \centering
  \includegraphics[width=0.6\linewidth]{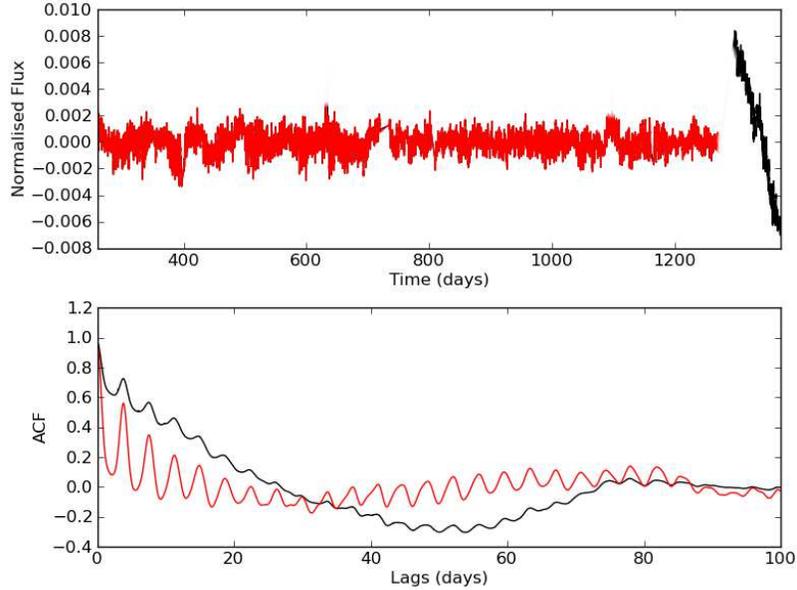}
  \caption{Example LC and ACF (KIC\,7286309), where the PDC-MAP correction of a quarter has failed, resulting in a steep slope at short lags in the ACF. The original LC and ACF are shown in black and the corrected ones in red. After removal of the faulty data, the ACF displays clearer peaks with higher LPHs.}
  \label{fig:corr_ex1}
\end{figure}

The following LC and ACF corrections were applied to those targets not initially selected as periodic by AutoACF. Large faults in the LC, such as discontinuities and uncorrected quarters, can cause a steep underlying slope in the ACF, and while the ACF peaks are still visible by eye, the LPH is reduced to a level not selected by AutoACF. We therefore performed an automatic search to remove sections of data containing faults or uncorrected data. This was done by rejecting quarters where the range of normalized flux between the $5^{\rm th}$ and $95^{\rm th}$ percentile was more than 5 times the median range of the other quarters. Running AutoACF on the fault-corrected LCs resulted in an additional 56 period detections, marked with the flag `BQR' in Table~\ref{tab:main_res}. An example LC and ACF before and after fault removal is shown are Figure~\ref{fig:corr_ex1}.

In cases where more than one period are present in the LC, the ACF displays an underlying oscillation from the longest period, with small peaks superimposed on it for the shorter periods. These short period peaks in the ACF typically have a low LPHs due to the underlying large peaks from the longer periods, which lead to a rejection of the detection by AutoACF. To deal with cases of multiple periods in the LC, we smoothed the ACF to allow detection of the larger underlying peaks, instead of the short period oscillations superimposed on it. The smoothing was performed by convolving the ACF with a Gaussian kernel, of window size of either 60 data points and full width at half-maximum (FWHM) of 20 data points (soft smoothing) or a window size of 200 data points and FWHM of 100 data points (hard smoothing). 

\begin{figure}
  \centering
  \includegraphics[width=0.6\linewidth]{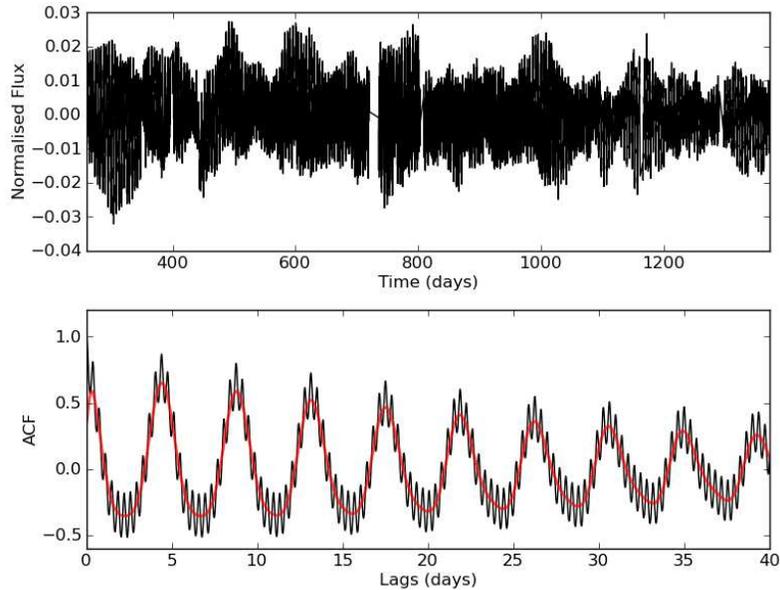}
  \caption{Example LC and ACF (KIC\,6064171), which displays both high- and low-frequency periodicity. The smoothed ACF (shown in red) has clear, high LPH peaks, which can be detected by AutoACF.}
  \label{fig:corr_ex2}
\end{figure}

We first applied the `soft smoothing' to all targets for which no period was previously detected, and visually inspected the results of AutoACF. This provided an additional 253 period detections, marked with the flag `SM1' in Table~\ref{tab:main_res}. We then ran the `hard smoothing' on the remaining targets, and those identified by eye as requiring further smoothing from the `soft smoothing' set. This process, followed by visual examination of the results, yielded an additional 125 rotation periods, marked with the flag `SM2' in Table~\ref{tab:main_res}. An example LC and ACF for a smoothed case are shown in Figure~\ref{fig:corr_ex2}. These additions bring the total number of rotation period detections to 34,030.

\end{document}